\documentclass[11pt,a4paper]{article}

\usepackage[utf8]{inputenc}
\usepackage{microtype}
\usepackage[margin=2.6cm]{geometry}
\usepackage{amsmath,amssymb}
\usepackage{tikz}
\usetikzlibrary{arrows.meta,calc}
\usepackage[colorlinks=true,linkcolor=blue!50!black,citecolor=blue!50!black,urlcolor=blue!50!black]{hyperref}

\numberwithin{equation}{section}

\newcommand{\Sig}{\Sigma}          
\newcommand{\qq}{q}                
\newcommand{\D}{D}                 
\newcommand{\Hs}{\mathcal{H}_0}    
\newcommand{\Hv}{\mathcal{H}}      
\newcommand{\Lie}{\pounds}         
\newcommand{\fR}{{}^{4}\!R}        
\newcommand{\sR}{{}^{3}\!R}        
\newcommand{\fG}{{}^{4}\Gamma}     
\newcommand{\sG}{{}^{3}\Gamma}     
\newcommand{\e}[2]{e_{#1}{}^{#2}}  
\newcommand{\ed}[2]{e^{#1}{}_{#2}} 
\newcommand{\pd}{\partial}
\newcommand{\dd}{\mathrm{d}}

\begin{document}

\begin{center}
{\LARGE\bfseries Introduction to the ADM Formalism}\\[1.2em]
{\large Alejandro Corichi$^{a,}$\footnote{\texttt{corichi@matmor.unam.mx}} \quad and \quad Darío Núñez$^{b}$}\\[0.6em]
{\small $^{a,b}$Instituto de Ciencias Nucleares, Universidad Nacional Autónoma de México,\\
A.\,Postal 70-543, 04510 México, D.\,F., México}\\[1.0em]
{\small Originally published in Spanish as \emph{Introducción al formalismo ADM},\\
Revista Mexicana de Física \textbf{37}, No.\,4 (1991) 720--747.}\\[0.4em]
{\small English translation by Ribhu Paul$^{c,}$\footnote{\texttt{ribhupaul.rp@gmail.com}, \texttt{psrp2452@iacs.res.in}} (arXiv:2210.10103)}\\[0.2em]
{\small $^{c}$Indian Association for the Cultivation of Science, 2A \& 2B Raja Subodh Chandra Mallick Rd,\\ Jadavpur, Kolkata 700032, India}\\[0.8em]
{\small Revised and expanded edition, October 2026}
\end{center}

\begin{abstract}
\noindent
We present the concepts and the mathematical developments that lead to the Hamiltonian formulation of Einstein's general relativity introduced by Arnowitt, Deser and Misner. All the geometric quantities needed for the construction are obtained explicitly, and some applications of the formalism are described. In this revised and expanded edition the original text has been rewritten in abstract index notation and corrected, and new sections have been added on the 3+1 form of Einstein's equations, the constraint algebra, boundary terms and the ADM energy, and on developments since 1991, together with a set of exercises.
\end{abstract}

\section{Introduction}

The formulation of general relativity established by Einstein is built around the principle of general covariance, according to which the laws of physics are independent of the state of motion of the observer. This principle implies that the laws must be written covariantly, that is, in a form independent of the coordinate system. The covariant formulation is therefore the natural one, and it expresses the interaction of matter with geometry in a simple and elegant way:
\begin{equation}
G_{ab} = \kappa\, T_{ab}\,,
\end{equation}
where $G_{ab}$ is the Einstein tensor and $T_{ab}$ is the energy--momentum tensor.

A feature of the covariance of Einstein's equations is that some of them are not evolution equations but constraints on the dynamical variables. These constraints are closely related to the contracted Bianchi identities,
\begin{equation}
\nabla^b G_{ab} = 0\,,
\end{equation}
which also guarantee that the constraints are preserved by the evolution whenever the dynamical equations are satisfied.

Despite the elegance of the covariant formulation of Einstein's equations, in practice it does not separate clearly the equations with dynamical content from those that are constraints.

One way to unravel the dynamics of general relativity is to regard it as a Cauchy problem, that is, to describe the dynamics as the evolution of a three-dimensional hypersurface on which the fields are defined. This reformulation of general relativity was developed by R.~Arnowitt, S.~Deser and C.\,W.~Misner; it is known as the ADM formulation of general relativity, and it reached its complete form in the early 1960s~\cite{ADM}.

The formalism has been developed continuously over its more than six decades of existence, although the interest of the community of theoretical physicists in its applications has risen and fallen over time. Since the mid-1980s there has been renewed and growing interest both in the formalism itself and in its applications to different areas. To mention a few examples: it served as the basis for the development of the Ashtekar variables~\cite{Ashtekar}; it has been used in the quantization of gravity by means of path integrals~\cite{GuvenRyan}; it is one of the main tools for solving Einstein's equations numerically~\cite{Matzner}; and it is at the basis of the theory of the space of three-geometries, or superspace~\cite{DeWitt}.

Given the importance of the formalism, and since there are, as far as we know, few clear and complete introductory expositions of it and of some of its applications, we have written this manuscript in an attempt to fill that gap. In these pages we give an introduction, as accessible as possible, to the ADM formalism as a Hamiltonian formulation of general relativity.

The paper is organized as follows. Section~\ref{sec:conceptual} presents, with minimal prerequisites, the ideas and concepts needed for the construction of the ADM formalism, in order to give a global view of the method. Section~\ref{sec:formal} contains the formal development, with all the equations written explicitly. Finally, in Sec.~\ref{sec:applications} we present some applications of the formalism and mention some of the areas that have been developed using it. In the present edition, Secs.~\ref{sec:einstein-3+1}--\ref{sec:ADM-energy} complete the formal development, Sec.~\ref{sec:developments} describes developments since 1991, and Appendix~\ref{app:exercises} collects exercises.

\paragraph{Note to this edition.}
This manuscript was written in 1991 and published in Spanish; an English translation of the original text appeared in 2022. The present edition has two components. Sections~1--4 (up to and including Sec.~3.7) are the original text, revised: we have kept its structure and its equation numbering, but have rewritten it in Penrose's abstract index notation~\cite{PenroseRindler}, following the conventions of Wald~\cite{Wald} (in particular for the sign of the extrinsic curvature, which is opposite to the one used in the original); we have improved the presentation and corrected a number of errors in the equations of the original, the most important being a missing factor of two in the total divergence of the action, the sign of the momentum constraint and of the shift term in the action, the expression of the Hamiltonian constraint, and the use of covariant instead of partial derivatives on minisuperspace. Sections~3.8--3.10, Sec.~5 and Appendix~A are new. They complete the formal development, with the 3+1 form of Einstein's equations, the analysis of the constraints and of the degrees of freedom, and the boundary terms that lead to the ADM energy, and they describe some of the developments since 1991 that build directly on the ADM formalism. Modern presentations of the 3+1 formalism can be found, for example, in~\cite{Wald,Gourgoulhon}. We hope that the paper may be useful to students interested in learning the 3+1 formalism of general relativity within the ADM framework.

\paragraph{Conventions.}
Spacetime is a four-dimensional manifold $M$ with a metric $g_{ab}$ of signature $(-,+,+,+)$. Latin letters $a,b,c,\dots$ from the beginning of the alphabet are abstract indices: they indicate the type of a tensor and do not take numerical values~\cite{PenroseRindler,Wald}. Greek letters $\mu,\nu,\ldots=0,\ldots,3$ denote components with respect to coordinates $x^\mu$ on spacetime, and Latin letters $i,j,k,\ldots=1,2,3$ from the middle of the alphabet denote components with respect to coordinates $\xi^i$ on a hypersurface. $\nabla_a$ is the derivative operator compatible with $g_{ab}$. The Riemann tensor is defined by $(\nabla_a\nabla_b-\nabla_b\nabla_a)\omega_c = R_{abc}{}^d\omega_d$, the Ricci tensor is $R_{ac}=R_{abc}{}^b$ and the scalar curvature is $R=g^{ac}R_{ac}$; with these conventions a round sphere has positive scalar curvature. Quantities intrinsic to a hypersurface carry a left superscript 3 (as in $\sR$) and, where confusion is possible, spacetime quantities carry a left superscript 4. We use units in which the coupling constant $16\pi G$ in front of the gravitational action is equal to one.

\section{Conceptual Development}\label{sec:conceptual}

In this section we explain in broad strokes the steps that lead to the ADM formalism. The purpose is to give a reasonably clear idea of the procedure before going into the details.

As mentioned in the introduction, the ADM variables should display explicitly the dynamical content of Einstein's equations and allow us to build a Hamiltonian all of whose terms have a reasonably clear interpretation. Once a Hamiltonian is defined, it is in principle possible to proceed to the quantization of the theory, after the true degrees of freedom have been isolated. In the ADM formulation, Einstein's equations are obtained from a variational principle.

The variational principle, one of the possible routes to Einstein's field equations, consists in finding the extrema of an action functional, in this case the gravitational action. The action must be expressed as the integral over spacetime, with the invariant volume element, of a scalar function.

In parallel with the work in which Einstein presented his theory of gravitation, Hilbert proposed a variational principle for it, with the action
\begin{equation}\label{eq:EH}
S=\int_M \dd^4x\,\sqrt{-g}\,R\,,
\end{equation}
where $g$ is the determinant of the components $g_{\mu\nu}$ of the metric and $R$ is the scalar curvature, so that the Lagrangian density is $\mathcal{L}=\sqrt{-g}\,R$.

The integral in~\eqref{eq:EH} extends over all of spacetime. The action is a functional of the metric, $S=S[g_{ab}]$, and the physically acceptable metrics are those for which the action is stationary; in other words, the action must be varied with respect to the metric in order to obtain the equations of motion. Performing this variation (taking due care with the boundary terms) and setting it equal to zero, we arrive at Einstein's equations in vacuum,
\begin{equation}
R_{ab}-\tfrac12\,R\,g_{ab}=0\,.
\end{equation}

Building a classical Hamiltonian theory from this starting point requires defining momenta canonically conjugate to the dynamical variables (the dynamical components of the metric), as partial derivatives of the Lagrangian density with respect to the generalized velocities. But how are these velocities to be defined? It is clear that one coordinate, time, must be singled out in order to define them. This breaks manifest covariance, and slices spacetime into hypersurfaces $x^0=\text{const}$. The algebraic complications that arise are considerable: P.\,A.\,M.~Dirac devoted a large part of the 1950s to this problem, and his Hamiltonian formulation of general relativity~\cite{Dirac58} appeared only shortly before the work of ADM.

It is therefore clear that, in order to construct a Hamiltonian, some coordinates must be separated from the others or, alternatively, a direction in spacetime must be singled out. One way of doing this is to consider slices of spacetime such that each slice is a three-dimensional hypersurface with a positive-definite metric. If we regard spacetime as formed by a family of such non-intersecting slices, each labelled by a number $t$ (which is not necessarily a time; it is just a label), we can think of evolution as the change of these hypersurfaces with the parameter $t$, as they sweep out all of spacetime. Each hypersurface inherits a three-dimensional metric $\qq_{ab}$, determined by the way spacetime has been sliced, and it is possible to regard this metric $\qq_{ab}(t)$, evolving in $t$, as the dynamical variable.

In addition to the six components of $\qq_{ab}$, four more variables are needed to make up the ten components of the original metric $g_{ab}$. These four variables arise naturally from the foliation by hypersurfaces. They are the \emph{lapse function} $N$, related to the separation between neighbouring hypersurfaces, and the \emph{shift vector} $N^a$, related to how a point is displaced when one passes to the next hypersurface. These four functions describe how the hypersurfaces are glued together to form the foliation.

Next, one rewrites the line element in terms of the new variables, $\dd s^2=\dd s^2[\qq_{ab},N,N^a]$, thereby relating the new variables to the old ones.

The following step is to rewrite the gravitational action in terms of the new variables. That is, we must express the spacetime scalar curvature $\fR$ and the invariant volume element $\dd^4x\sqrt{-g}$ in terms of geometric objects on the hypersurface and of $N$ and $N^a$. This requires studying how a hypersurface is embedded in spacetime, and how it inherits geometric structure both from spacetime and from the particular embedding (that is, from the way the slicing is done). It also requires studying how spacetime tensors are projected onto the hypersurface and onto the direction orthogonal to it.

In this step the extrinsic curvature $K_{ab}$ of the hypersurface plays a central role. It depends on the lapse, the shift and the velocity of the metric, $K_{ab}=K_{ab}[N,\D_aN_b,\dot\qq_{ab}]$, where the dot denotes the derivative with respect to the parameter $t$ and $\D_a$ is the covariant derivative on the hypersurface. With its help the Riemann tensor, and therefore the scalar curvature $\fR$, can be rewritten as a function of $\qq_{ab}$, $\dot \qq_{ab}$, $N$ and $N^a$. We emphasize that the new Lagrangian density will not depend on the derivatives of the lapse and the shift with respect to $t$. Finally, the action takes the form
\begin{equation}\label{eq:S-ADM-intro}
S[\qq_{ab},N,N^a]=\int \dd t\int \dd^3x\,\sqrt{\qq}\,N\big(K_{ab}K^{ab}-K^2+\sR+2\nabla_a\Delta^a\big)\,,
\end{equation}
where $\sqrt{-g}=\sqrt{\qq}\,N$ is the invariant volume element written in the new variables, $\qq$ is the determinant of the components of $\qq_{ab}$, and the last term is a total divergence whose explicit form is given in Sec.~\ref{sec:curvature-scalar}.

Since the Lagrangian density does not depend on the velocities of $N$ and $N^a$, these can be regarded as non-dynamical variables. This also suggests that the true dynamical variables are the six components of $\qq_{ab}(t)$.

In this way we obtain a reformulation of the variational principle of general relativity with a clearer dynamical content, which was our aim. We can now go further and construct the Hamiltonian density $\mathcal{H}$ from the new Lagrangian density in the usual way, by defining momenta canonically conjugate to $\qq_{ab}$,
\begin{equation}
\pi^{ab}=\frac{\pd\mathcal{L}}{\pd\dot \qq_{ab}}\,,
\end{equation}
and performing the Legendre transformation only on the dynamical variables and their momenta:
\begin{equation}
\mathcal{H}=\pi^{ab}\dot \qq_{ab}-\mathcal{L}\,.
\end{equation}
Up to a total divergence, this Hamiltonian density has the form
\begin{equation}
\mathcal{H}=N\Hs+N^a\Hv_a\,.
\end{equation}
The scalar density $\Hs$ and the covector density $\Hv_a$ will play a very important role in the analysis of the dynamics of the theory.

As mentioned at the beginning of this section, in general relativity there are constraints among the variables, a consequence of the covariance of the theory. With the ADM formalism, the form of these constraints and the variables involved in them become clear. First, there are four primary constraints, which are a direct consequence of the form of the Lagrangian density: the momenta conjugate to the four variables $N$ and $N^a$,
\begin{equation}
P=\frac{\pd\mathcal{L}}{\pd\dot N}\,,\qquad P_a=\frac{\pd\mathcal{L}}{\pd\dot N^a}\,,
\end{equation}
vanish weakly, since $\mathcal{L}$ depends neither on $\dot N$ nor on $\dot N^a$.

Using the Hamiltonian density we can now rewrite the Lagrangian density in terms of the dynamical variables $\qq_{ab}$ and their conjugate momenta $\pi^{ab}$, thus obtaining a first-order variational principle. The action is then written as
\begin{equation}\label{eq:S-first-order-intro}
S[\qq_{ab},\pi^{ab},N,N^a]=\int \dd t\int \dd^3x\,\big(\pi^{ab}\dot \qq_{ab}-N\Hs-N^a\Hv_a\big)\,.
\end{equation}
This expression of the action is in \emph{parametrized} form, as a consequence of covariance (see Sec.~\ref{sec:hamiltonian}). The field equations are obtained by varying with respect to the different variables. Varying with respect to $N$ and $N^a$ one finds $\Hs=0$ and $\Hv_a=0$: four relations among the dynamical variables $\qq_{ab}$ and their momenta $\pi^{ab}$. These are the constraints of the theory; $\Hs=0$ is known as the Hamiltonian (or scalar) constraint and $\Hv_a=0$ as the momentum (or vector, or diffeomorphism) constraint. This completes our presentation of the fundamental ideas behind the construction of the ADM formalism as a Hamiltonian formulation. In the next section we carry out the construction explicitly.

\section{Formal Development}\label{sec:formal}

\subsection{Embedding a Hypersurface in a Manifold}

Let spacetime be a four-dimensional manifold $M$ with a metric $g_{ab}$ of signature $(-,+,+,+)$, and let $x^\mu$ be coordinates on $M$. Let $\Sig$ be a three-dimensional manifold with coordinates $\xi^i$. An embedding of $\Sig$ in $M$ is a map $X:\Sig\to M$, given in coordinates by
\begin{equation}
x^\mu=X^\mu(\xi^i)\,,
\end{equation}
where $\mu=0,1,2,3$ and $i=1,2,3$. These four functions determine the embedding. For $X$ to be an embedding the hypersurface must not intersect itself; more precisely, $X$ must be an injective immersion and a homeomorphism onto its image~\cite{Spivak}. We use the same symbol $\Sig$ for the image $X(\Sig)\subset M$.

The metric on $M$ induces a metric on $\Sig$ when the line element is restricted to $\Sig$:
\begin{equation}
\begin{split}
\dd s^2 &= g_{\mu\nu}(x)\,\dd x^\mu \dd x^\nu\Big|_{\Sig}
= g_{\mu\nu}(X)\,\frac{\pd X^\mu}{\pd\xi^i}\,\dd\xi^i\,\frac{\pd X^\nu}{\pd\xi^j}\,\dd\xi^j\\
&= \Big(g_{\mu\nu}(X)\,\frac{\pd X^\mu}{\pd\xi^i}\frac{\pd X^\nu}{\pd\xi^j}\Big)\dd\xi^i \dd\xi^j
\equiv \qq_{ij}\,\dd\xi^i \dd\xi^j\,,
\end{split}
\end{equation}
so that the induced metric has components $\qq_{ij}=g_{\mu\nu}(X)\,\frac{\pd X^\mu}{\pd\xi^i}\frac{\pd X^\nu}{\pd\xi^j}$. Let us define
\begin{equation}
X^\mu{}_i\equiv\frac{\pd X^\mu}{\pd\xi^i}\,;
\end{equation}
$X^\mu{}_i$ is the $\mu$-th component (along the coordinate $x^\mu$) of the $i$-th vector of the coordinate basis of $\Sig$,
\begin{equation}
\e{i}{a}\equiv\Big(\frac{\pd}{\pd\xi^i}\Big)^a\,,
\end{equation}
pushed forward to $M$ by the embedding. Indeed, regarded as a vector in $M$ tangent to $\Sig$, $\e{i}{a}$ is
\begin{equation}
\e{i}{a}=X^\mu{}_i\Big(\frac{\pd}{\pd x^\mu}\Big)^a\,,\qquad\text{i.e.}\qquad
\frac{\pd X^\mu}{\pd\xi^i}\frac{\pd}{\pd X^\mu}=\frac{\pd}{\pd\xi^i}\,.
\end{equation}
We then have a natural expression for the components of the metric on $\Sig$,
\begin{equation}
\qq_{ij}=g_{ab}\,\e{i}{a}\e{j}{b}\,.
\end{equation}
For the study of the dynamics of spacetime the embedded hypersurface must be spacelike, so we require the metric $\qq_{ij}$ to be positive definite, with signature $(+,+,+)$.

The three vectors $\e{i}{a}$ form a basis of the tangent space $T_p\Sig$ at each point $p$ of $\Sig$. This space is a subspace of the tangent space $T_pM$. To complete a basis of $T_pM$ we construct the orthogonal complement of $T_p\Sig$ with respect to $g_{ab}$. It is spanned by a single vector, which we denote by $n^a$; it satisfies
\begin{equation}
g_{ab}\,\e{i}{a}n^b = n_a\,\e{i}{a}=0\,,
\end{equation}
and we impose the normalization
\begin{equation}
g_{ab}\,n^an^b=-1\,.
\end{equation}
These two conditions determine $n^a$ up to sign; we choose it to point to the future.

The set $(\e{i}{a},n^a)$ is then a basis of $T_pM$ at each point $p$ of $\Sig$. With it, any vector $v^a$ of $T_pM$ can be written as a linear combination of the basis vectors of $T_p\Sig$ and the normal:
\begin{equation}\label{eq:decomposition}
v^a=v^i\,\e{i}{a}+v^\perp n^a\,,
\end{equation}
where
\begin{equation}\label{eq:vperp}
v^\perp=-n_bv^b\,.
\end{equation}
If $v^a$ is tangent to the hypersurface, it has the form $v^a=v^i\,\e{i}{a}$.

The coefficients $v^i$ behave as scalars under changes of the spacetime coordinates $x^\mu\to x'^\mu$, since under such changes neither the vectors $\e{i}{a}$ nor $v^a$ change. They are, however, the components of a vector of $T_p\Sig$, since they transform as such under changes $\xi^i\to\xi'^i$ of the coordinates on the hypersurface:
\begin{equation}
v^i\Big(\frac{\pd}{\pd\xi^i}\Big)^a=v^i\,\frac{\pd\xi'^j}{\pd\xi^i}\Big(\frac{\pd}{\pd\xi'^j}\Big)^a=v'^j\,e'_j{}^a\,,
\end{equation}
so that the new components are
\begin{equation}
v'^j=\frac{\pd\xi'^j}{\pd\xi^i}\,v^i\,.
\end{equation}

We define the inverse metric $\qq^{ij}$ on $\Sig$ by $\qq^{ij}\qq_{jk}=\delta^i{}_k$, and we use $\qq_{ij}$ and $\qq^{ij}$ to lower and raise the indices $i,j,\dots$:
\begin{equation}
v_i=\qq_{ij}v^j\,,\qquad v^i=\qq^{ij}v_j\,.
\end{equation}
Writing $v_j\equiv g_{ab}\,\e{j}{a}v^b$, the components of a vector along the hypersurface can be expressed as
\begin{equation}
v^i=\qq^{ij}v_j=\qq^{ij}g_{ab}\,\e{j}{a}v^b=\ed{i}{b}v^b\,,
\end{equation}
where we have defined the covectors
\begin{equation}
\ed{i}{b}\equiv \qq^{ij}g_{ab}\,\e{j}{a}\,,
\end{equation}
which extract the $i$-th component of any vector. They satisfy $\ed{i}{a}\e{j}{a}=\delta^i{}_j$ and $\ed{i}{a}n^a=0$.

Given an arbitrary spacetime vector $v^a$, we now know how to find its projections onto the hypersurface, $v_i=g_{ab}\,\e{i}{a}v^b$, in the natural basis of $T_p\Sig$. We now want an expression for the same projection as a spacetime vector, without reference to the basis. To this end, note that
\begin{equation}
v^a=v^i\,\e{i}{a}-(n_bv^b)\,n^a\,.
\end{equation}
The projection of $v^a$ onto the hypersurface is therefore
\begin{equation}
v^i\,\e{i}{a}=v^a+(n_bv^b)\,n^a=(\delta^a{}_b+n^an_b)\,v^b\,.
\end{equation}
We have thus found the operator that projects spacetime vectors onto the hypersurface. We denote it by
\begin{equation}\label{eq:projector}
\qq^a{}_b\equiv\delta^a{}_b+n^an_b=\e{i}{a}\,\ed{i}{b}\,.
\end{equation}
Using (3.7) and (3.8) it is easy to check that it satisfies
\begin{equation}
\qq^a{}_b\,n^b=0\,,\qquad \qq^a{}_b\,\qq^b{}_c=\qq^a{}_c\,,\qquad \qq^a{}_a=3\,.
\end{equation}

Lowering its index with $g_{ab}$ we obtain the tensor $\qq_{ab}=g_{ab}+n_an_b$, whose components in the basis $\e{i}{a}$ are precisely those of the induced metric, $\qq_{ab}\,\e{i}{a}\e{j}{b}=\qq_{ij}$. The induced metric, so far a tensor on $\Sig$, can thus be regarded as a spacetime tensor orthogonal to $n^a$. More generally, the vectors $\e{i}{a}$ and covectors $\ed{i}{a}$ identify tensors on $\Sig$ with spacetime tensors all of whose contractions with $n^a$ and $n_a$ vanish; we call such tensors \emph{spatial}. From now on we make this identification, which allows us to use the same abstract indices $a,b,c,\dots$ for tensors on $\Sig$ and on $M$: $\qq_{ab}$ denotes at the same time the induced metric on $\Sig$ and the spatial tensor $g_{ab}+n_an_b$, and $\qq^{ab}=g^{ab}+n^an^b$ is its inverse on spatial tensors. The components of a spatial tensor are obtained by contracting with the basis, as in $v^i=\ed{i}{a}v^a$ or $\qq_{ij}=\e{i}{a}\e{j}{b}\qq_{ab}$.

\subsection{Induced Covariant Derivatives}

The covariant derivative defined by the metric $g_{ab}$ of a vector field tangent to $\Sig$, taken in the direction of another vector tangent to $\Sig$, is in general not tangent to the hypersurface. The natural way to define a covariant derivative on $\Sig$ is to require that it give again vectors of $T_p\Sig$~\cite{MTW}. At this point it is convenient to consider coordinates in which three of the $x^\mu$ coincide with the $\xi^i$, say $x^i=\xi^i$, while $x^0$ is constant on $\Sig$. These are called \emph{adapted coordinates}. In them the vectors $\e{i}{a}$ coincide with three of the vectors of the coordinate basis of $T_pM$, so that their components are trivial: $X^\mu{}_i=\delta^\mu{}_i$. Using these coordinates generally simplifies the analysis without loss of generality, since it is always possible to take the three intrinsic coordinates of $\Sig$ as three of the spacetime coordinates. From now on, components will be taken in adapted coordinates. Consider then the spacetime covariant derivative of a vector $v^a=v^j\,\e{j}{a}$ tangent to the hypersurface, in the direction of a basis vector:
\begin{equation}
\begin{split}
\e{i}{b}\nabla_b v^a &=\e{i}{b}\nabla_b\big(v^j\,\e{j}{a}\big)
=\e{j}{a}\,\frac{\pd v^j}{\pd\xi^i}+v^j\,\fG^\mu{}_{ji}\Big(\frac{\pd}{\pd x^\mu}\Big)^a\\
&=\e{j}{a}\,\frac{\pd v^j}{\pd\xi^i}+v^j\Big(\fG^k{}_{ji}\,\e{k}{a}+\fG^0{}_{ji}\Big(\frac{\pd}{\pd x^0}\Big)^a\Big)\,,
\end{split}
\end{equation}
where $\fG^\mu{}_{\nu\rho}$ are the Christoffel symbols of $g_{ab}$ in the coordinates $x^\mu$. As a special case, the covariant derivative of the basis vectors is
\begin{equation}
\e{i}{b}\nabla_b\e{j}{a}=\fG^\mu{}_{ji}\Big(\frac{\pd}{\pd x^\mu}\Big)^a=\fG^k{}_{ji}\,\e{k}{a}+\fG^0{}_{ji}\Big(\frac{\pd}{\pd x^0}\Big)^a\,.
\end{equation}
In both (3.20) and (3.21) there is a term that points out of the hypersurface. Its normal component,
\begin{equation}
-\big(\fG^0{}_{ji}\,v^j\big)\Big(n_a\Big(\frac{\pd}{\pd x^0}\Big)^a\Big)\,,
\end{equation}
does not vanish in general. We can get rid of it by projecting the vector onto the hypersurface, so as to obtain a vector of $T_p\Sig$. In this way we obtain the covariant derivative intrinsic to the three-geometry of the hypersurface, which we denote by $\D_a$. For a vector $v^a=v^j\,\e{j}{a}$ on the hypersurface, the components of its covariant derivative are
\begin{equation}
\D_iv^k\equiv\e{i}{a}\,\ed{k}{b}\,\D_av^b=\pd_iv^k+\sG^k{}_{ji}\,v^j\,,
\end{equation}
where $\sG^k{}_{ji}=\qq^{kl}\,\sG_{lji}$, and, as we show below, $\sG_{lji}=\fG_{lji}$: the Christoffel symbols with all indices down coincide. The corresponding statement with the index up is false in general. Raising the index with the spacetime inverse metric, $\fG^k{}_{ji}=g^{k\mu}\fG_{\mu ji}$ picks up a contribution from $\fG_{0ji}$ whenever $g^{k0}\neq0$, that is, whenever $(\pd/\pd x^0)^a$ is not orthogonal to $\Sig$; we shall see in Sec.~\ref{sec:3+1} that $\fG^k{}_{ji}=\sG^k{}_{ji}-(N^k/N)K_{ij}$.

If we now consider the $k$-th component of the covariant derivative of $v^a$ in the direction $\e{i}{a}$, with the index down, it takes the form
\begin{equation}
\D_iv_k\equiv g_{ab}\,\e{k}{a}\,\e{i}{c}\nabla_cv^b\,,
\end{equation}
which can also be expressed as
\begin{equation}
\D_iv_k=\pd_iv_k-\sG^j{}_{ki}\,v_j\,.
\end{equation}
Note that the covariant derivative takes the same form, both for the components of a vector and for those of a covector, as the covariant derivative on spacetime. This is because the covariant derivative on the hypersurface is given by the affine connection of $\Sig$ defined in the usual way in terms of its metric (\cite{DeWitt} and~\cite{CorichiThesis}),
\begin{equation}
\sG_{kij}\equiv\tfrac12\big(\pd_j\qq_{ki}+\pd_i\qq_{kj}-\pd_k\qq_{ij}\big)\,,
\end{equation}
with the interpretation
\begin{equation}
\sG_{kji}=g_{ab}\,\e{k}{a}\,\e{i}{c}\nabla_c\e{j}{b}\,.
\end{equation}
Indeed, in adapted coordinates the right-hand side of (3.27) is $g_{k\mu}\fG^\mu{}_{ji}=\fG_{kji}$, which involves only the derivatives along $\Sig$ of $g_{jk}=\qq_{jk}$ (see (3.6)) and therefore coincides with (3.26). $\D_a$ is thus the unique torsion-free derivative operator on $\Sig$ compatible with its metric, $\D_a\qq_{bc}=0$.

In practice it is convenient to note that, for $v^a$ tangent to $\Sig$, the covariant derivative can be written without reference to any basis as
\begin{equation}\label{eq:D-projection}
\D_cv^b=\qq^b{}_a\,\qq_c{}^d\,\nabla_dv^a\,.
\end{equation}
This is the most practical expression, and the one we shall use from now on; we have nevertheless preferred to arrive at it through components, since this makes its geometrical meaning clearer. It extends to spatial tensors of any type by projecting every index, $\D_c T^{a\dots}{}_{b\dots}=\qq_c{}^d\,\qq^a{}_e\cdots \qq_b{}^f\cdots\nabla_dT^{e\dots}{}_{f\dots}$.

\subsection{Extrinsic Curvature}

We now define the extrinsic curvature of the hypersurface as\footnote{This is the sign convention of Wald~\cite{Wald}. The original version of this paper used the opposite sign, as in~\cite{MTW}. With the present convention $K=\qq^{ab}K_{ab}=\nabla_an^a$ is positive when the normals diverge, as for the hypersurfaces of homogeneity of an expanding universe.}
\begin{equation}\label{eq:K-def}
K_{ab}\equiv \qq_a{}^c\,\qq_b{}^d\,\nabla_cn_d\,,
\end{equation}
that is, the component $K_{ij}=K_{ab}\,\e{i}{a}\e{j}{b}$ of the extrinsic curvature is the projection along $\e{j}{a}$ of the covariant derivative of the normal in the direction $\e{i}{a}$. The extrinsic curvature is well defined as a tensor on the hypersurface, since the vector $\qq_a{}^c\nabla_cn^b$ is tangent to $\Sig$: from
\[
\nabla_c(n^bn_b)=0
\]
it follows that
\begin{equation}
n^b\nabla_cn_b=0\,,
\end{equation}
so that $\qq_b{}^d\nabla_cn_d=\nabla_cn_b$ and the definition can be written more simply as $K_{ab}=\qq_a{}^c\nabla_cn_b$, or as
\[
\qq_a{}^c\,\nabla_cn^b=K_a{}^b\,.
\]
$K_{ab}$ is thus a geometric object on $\Sig$. The notion of extrinsic curvature has no meaning for a manifold by itself; it acquires a meaning only when the manifold is embedded in another of higher dimension since, by its very definition, $K_{ab}$ depends on the geometry of the larger manifold (through the spacetime covariant derivative of the normal to the hypersurface).

A geometric interpretation of the extrinsic curvature is that it measures how much the hypersurface is curved within $M$; in other words, it tells us how the normals at two nearby points of $\Sig$ fail to be parallel (see Fig.~\ref{fig:K}). There is another way of writing the extrinsic curvature, which follows from the orthogonality of $n^a$ to the hypersurface, $n_av^a=0$ for every vector $v^a$ tangent to $\Sig$. For vector fields $v^a$ and $w^a$ tangent to $\Sig$ we then have
\begin{equation}\label{eq:K-alt}
K_{ab}\,v^aw^b=-\,n_b\,v^c\nabla_cw^b\,.
\end{equation}
When $v^a$ and $w^a$ are exchanged, the right-hand side changes by $n_b[v,w]^b$, which vanishes because the commutator of two vector fields tangent to $\Sig$ is again tangent to $\Sig$. Hence the extrinsic curvature is symmetric, $K_{ab}=K_{ba}$.

\begin{figure}[t]
\centering
\begin{tikzpicture}[>=Stealth, scale=1.05]
  \def\R{4}
  \coordinate (O) at (0,-3);
  \draw[thick] ($(O)+(126:\R)$) arc (126:54:\R);
  \node at ($(O)+(54:\R)+(0.35,0.1)$) {$\Sigma$};
  \coordinate (p)  at ($(O)+(100:\R)$);
  \coordinate (pp) at ($(O)+(80:\R)$);
  \fill (p) circle (1.5pt) node[below left=1pt] {$p$};
  \fill (pp) circle (1.5pt) node[below right=1pt] {$p'$};
  \draw[->,thick] (p) -- ($(p)+(100:2.6)$) node[left=2pt] {$n^a(p)$};
  \draw[->,thick] (pp) -- ($(pp)+(80:2.6)$) node[above right] {$n^a(p')$};
  \draw[->,dashed] (p) -- ($(p)+(80:2.6)$);
  \draw[->,very thick,gray!70!black] ($(p)+(100:2.6)$) -- ($(p)+(80:2.6)$)
     node[midway,above=3pt,black] {$\delta n^b$};
  \draw[->,gray] ($(O)+(98.5:\R)+(0,-0.25)$) arc (98.5:81.5:\R-0.25)
     node[midway,below=1pt,black] {$\delta x^a$};
\end{tikzpicture}
\caption{Schematic representation of the extrinsic curvature. The normal at $p'$, transported back to $p$ (dashed), differs from the normal at $p$ by $\delta n^b=K_a{}^b\,\delta x^a$, where $\delta x^a$ is the displacement from $p$ to $p'$ along $\Sigma$.}
\label{fig:K}
\end{figure}

This last expression and Eq.~(3.27) tell us how to write the vector $\e{i}{b}\nabla_b\e{j}{a}$ in terms of its components along the hypersurface and along the normal:
\begin{equation}\label{eq:Gauss-formula}
\e{i}{b}\nabla_b\e{j}{a}=\sG^k{}_{ji}\,\e{k}{a}+K_{ij}\,n^a\,.
\end{equation}
This relation is known as the \emph{Gauss formula}. Knowing the spacetime covariant derivative of the basis vectors, we can write the derivative of an arbitrary vector field $v^a$ tangent to $\Sig$:
\begin{equation}
\qq_c{}^b\,\nabla_bv^a=\D_cv^a+K_{cb}\,v^b\,n^a\,.
\end{equation}
To finish this section, we write the extrinsic curvature in components,
\begin{equation}
K_{ij}=X^\mu{}_i\,X^\nu{}_j\,\nabla_\mu n_\nu\,,
\end{equation}
so that in adapted coordinates it takes the form
\begin{equation}
K_{ij}=\nabla_in_j\,,
\end{equation}
where $\nabla_\mu n_\nu$ are the components of $\nabla_an_b$.

\subsection{Decomposition of Spacetime into 3+1}\label{sec:3+1}

So far we have described precisely how a three-geometry is embedded in spacetime: we have introduced the metric $\qq_{ab}$ on the hypersurface and the covariant derivative $\D_a$ defined on it (and hence the notion of parallel transport on $\Sig$). We also know precisely how to decompose a spacetime vector into its components tangent and orthogonal to the hypersurface.

The next step is to require that the whole of spacetime be generated by such hypersurfaces (each corresponding to an embedding) without them intersecting each other. The family of these hypersurfaces is called a \emph{foliation}. As an example, Euclidean space $\mathbb{R}^3$ with the origin removed can be seen as generated by spheres of radius $a$: each sphere is a hypersurface (of dimension 2) embedded in $\mathbb{R}^3$ (of dimension 3), and the family of all these spheres is a foliation of $\mathbb{R}^3\setminus\{0\}$.

The foliation is given by a function $t$ on spacetime such that each hypersurface corresponds to a constant value of $t$, i.e.\ the hypersurfaces are the level sets $\Sig_t=\{t=\text{const}\}$. (Recall that $t$ is simply a label.) In the example of the spheres, such a function can be the radius $r=r(x,y,z)$, which takes the value $r=a$ on each sphere.

\begin{figure}[t]
\centering
\begin{tikzpicture}[>=Stealth, scale=0.95]
  \shade[ball color=gray!8, opacity=0.9] (0,0) circle (2.7);
  \draw (0,0) circle (2.7);
  \draw[gray, densely dashed] (-2.7,0) arc (180:360:2.7 and 0.75);
  \shade[ball color=gray!35] (0,0) circle (1.6);
  \draw (0,0) circle (1.6);
  \draw[gray] (-1.6,0) arc (180:360:1.6 and 0.45);
  \draw[->] (0,0) -- (0,3.5) node[above] {$z$};
  \draw[->] (0,0) -- (3.6,0) node[right] {$y$};
  \draw[->] (0,0) -- (-2.3,-2.3) node[below left] {$x$};
  \coordinate (p) at (0,1.6);
  \coordinate (pp) at (72:2.7);
  \fill (p) circle (1.6pt) node[left=2pt] {$p$};
  \fill (pp) circle (1.6pt) node[above right=0pt] {$p'$};
  \draw (0,2.7) circle (1.6pt);
  \draw[->,thick] (p) -- (pp);
  \node[right] at (2.15,-1.85) {$\Sigma_{t+\dd t}:\ r=a+\dd r$};
  \draw (1.95,-1.7) -- (2.2,-1.85);
  \node[right] at (1.7,-2.6) {$\Sigma_{t}:\ r=a$};
  \draw (1.0,-1.2) -- (1.7,-2.6);
\end{tikzpicture}
\caption{Two spheres of the foliation of $\mathbb{R}^3\setminus\{0\}$. The points $p$ and $p'$ have the same intrinsic coordinate $\theta=0$ on their respective spheres. The point $p'$ need not be the north pole of the outer sphere (open circle), so the vector joining $p$ to $p'$ is in general not orthogonal to the inner sphere.}
\label{fig:spheres}
\end{figure}

Given the hypersurface with label $t=t_0$, a point $p$ on it is characterized by its three coordinates $\xi^i$; the point represents an event. To it corresponds a point $p'$ on the hypersurface with label $t_0+\dd t$, namely the point with the same intrinsic coordinates. Where $p'$ lies with respect to the hypersurface $t_0$ depends on how the intrinsic coordinates are chosen on each hypersurface: $p'$ has, on the hypersurface labelled $t_0+\dd t$, the same intrinsic coordinates as $p$ on the original hypersurface, but its spacetime coordinates are different, since they depend on $t$ as well.

To illustrate this idea, take as $p$ the north pole of the sphere of radius $a$ (Fig.~\ref{fig:spheres}); it has the coordinate $\theta=0$ on the sphere. The point $p'$ on the sphere of radius $a+\dd r$ that corresponds to $p$ need not be the north pole of that sphere (the point on the $z$ axis of $\mathbb{R}^3$), even though it has the same intrinsic coordinate $\theta=0$ on its own sphere. There is then a vector joining $p$ and $p'$, based at $p$ (we take $\dd t$ infinitesimal), which in general is not orthogonal to the hypersurface. For the two spheres of the figure, the vector is orthogonal only if $p'$ is also the north pole.

Analytically, the foliation is given by the functions
\begin{equation}
x^\mu=X^\mu(\xi^i,t)\,,
\end{equation}
and the vector joining points of neighbouring hypersurfaces is the \emph{evolution vector field} $t^a=(\pd/\pd t)^a$, whose components in the basis $(\pd/\pd x^\mu)^a$ are
\begin{equation}
t^a=\Big(\frac{\pd}{\pd t}\Big)^a=\frac{\pd X^\mu}{\pd t}\Big(\frac{\pd}{\pd x^\mu}\Big)^a\,;
\end{equation}
it satisfies $t^a\nabla_at=1$. From Sec.~3.1 we know how to decompose a vector into its projection onto the hypersurface and its normal component, Eqs.~\eqref{eq:decomposition} and~\eqref{eq:vperp}, so we can write
\begin{equation}\label{eq:lapse-shift}
t^a=-(n_bt^b)\,n^a+\qq^a{}_b\,t^b\equiv N n^a+N^a\,.
\end{equation}
The scalar $N=-n_at^a$ is called the \emph{lapse function}, and the spatial vector $N^a=\qq^a{}_bt^b$ is called the \emph{shift vector}. Together with the metric $\qq_{ab}$ they constitute the ADM variables. Since $n_a$ is orthogonal to the level sets of $t$ and $t^a\nabla_at=1$, the normal and the lapse are related by $n_a=-N\nabla_at$.

As the example of the spheres shows, the lapse function tells us how far apart the next sphere is; in that case the separation is the same at every point, since all the hypersurfaces are spheres, but in general it varies from point to point. The shift vector, in turn, tells us how far the point with the same intrinsic coordinates has slid along the sphere, that is, how much the coordinates have been rotated or, more generally, deformed. The interpretation for spacelike hypersurfaces is analogous, and the names of the new variables $N$ and $N^a$ are therefore appropriate. The lapse function gives the lapse of time between the events $p$ and $p'$ (for an observer moving along the normal, $N\,\dd t$ is the proper time elapsed), since the direction normal to the hypersurface is timelike (recall that spacetime has signature $(-,+,+,+)$). The shift vector tells us how much the point with fixed $\xi^i$ is displaced along the hypersurface and therefore, seen globally, how the coordinates on the hypersurfaces are deformed.

It is clear from Eq.~\eqref{eq:lapse-shift} that, given a hypersurface, specifying the four functions $N$ and $N^a$ determines how the next hypersurface of the foliation is placed (the lapse) and how the spatial coordinates are carried from one hypersurface to the next (the shift). What is not determined in this way is the metric induced on each hypersurface; it will be given by the dynamical equations derived below.

Let us now rewrite the spacetime line element (i.e.\ the interval between two events) in terms of the new variables $\qq_{ab}$, $N$ and $N^a$. The interval between the events $p(x^i,t)$ and $p'(x^i+\dd x^i,t+\dd t)$ can be decomposed into two parts: the square of the distance along the hypersurface containing $p$, minus the square of the proper time between the hypersurfaces,
\begin{equation}
\begin{split}
\dd s^2&=\qq_{ij}\,(\dd x^i+N^i\dd t)(\dd x^j+N^j\dd t)-(N\dd t)^2\\
&=(N^iN_i-N^2)\,\dd t^2+2N_i\,\dd x^i\dd t+\qq_{ij}\,\dd x^i\dd x^j\,,
\end{split}
\end{equation}
where $(\dd x^i+N^i\dd t)$ is the displacement along the base hypersurface and $N\dd t$ is the proper time between the hypersurfaces (see Fig.~\ref{fig:lapse-shift}). This can be seen most clearly in the case $N^i=0$: the line element then has contributions from (i) the distance along the base hypersurface and (ii) the proper time, since the evolution of the hypersurfaces is then purely temporal ($t^a$ is normal to the hypersurfaces). In abstract index notation, (3.39) is simply the statement $g_{ab}=\qq_{ab}-n_an_b$ with $n_a=-N\nabla_at$.

\begin{figure}[t]
\centering
\begin{tikzpicture}[>=Stealth, scale=1.0]
  \draw[thick, fill=gray!10] (0,0) .. controls (2,0.55) and (4,-0.35) .. (6,0.3)
     -- (7.6,2.0) .. controls (5.6,1.4) and (3.6,2.25) .. (1.6,1.75) -- cycle;
  \node at (6.55,0.45) {$\Sigma_t$};
  \begin{scope}[yshift=2.7cm]
  \draw[thick, fill=gray!4, fill opacity=0.85] (0,0) .. controls (2,0.55) and (4,-0.35) .. (6,0.3)
     -- (7.6,2.0) .. controls (5.6,1.4) and (3.6,2.25) .. (1.6,1.75) -- cycle;
  \node at (6.9,0.55) {$\Sigma_{t+\dd t}$};
  \end{scope}
  \coordinate (p) at (3.2,0.95);
  \coordinate (q) at (3.2,3.65);
  \coordinate (s) at (4.6,1.30);
  \coordinate (pp) at (4.6,4.00);
  \draw[->,thick] (p) -- (q) node[midway,left] {$N n^a$};
  \draw[->,thick] (p) -- (s) node[midway,below right=-1pt] {$N^a$};
  \draw[->,very thick] (p) -- (pp) node[midway,right=3pt] {$t^a$};
  \draw[densely dashed] (q) -- (pp);
  \draw[densely dashed] (s) -- (pp);
  \fill (p) circle (1.6pt) node[below left=0pt] {$p$};
  \fill (pp) circle (1.6pt) node[above right=0pt] {$p'$};
  \node[anchor=west] at (-0.3,4.95) {$t^a=N n^a+N^a$};
\end{tikzpicture}
\caption{Decomposition of the evolution vector $t^a=(\pd/\pd t)^a$ into its component $N n^a$ along the normal and its projection $N^a$ onto $\Sigma_t$. The point $p'$ on $\Sigma_{t+\dd t}$ has the same intrinsic coordinates as $p$ on $\Sigma_t$.}
\label{fig:lapse-shift}
\end{figure}

From (3.39) we read off the relation between the components of the metric $g_{ab}$ in adapted coordinates $(t,x^i)$ and the variables $\qq_{ij}$, $N$ and $N^i$:
\begin{equation}
\begin{split}
g_{00}&=N^iN_i-N^2\,,\\
g_{0i}&=N_i\,,\\
g_{ij}&=\qq_{ij}\,,
\end{split}
\end{equation}
where $N_i=\qq_{ij}N^j$. Correspondingly, the components of the inverse metric are
\begin{equation}
g^{\mu\nu}=
\begin{pmatrix}
-\dfrac{1}{N^2} & \dfrac{N^j}{N^2}\\[2ex]
\dfrac{N^i}{N^2} & \qq^{ij}-\dfrac{N^iN^j}{N^2}
\end{pmatrix},
\end{equation}
as follows from $g^{ab}=\qq^{ab}-n^an^b$ with $n^a=(t^a-N^a)/N$; here $\qq^{ij}$ is the inverse of the $3\times3$ matrix $\qq_{ij}$, not the spatial part of $g^{\mu\nu}$. The volume element is then
\begin{equation}
\sqrt{-g}\;\dd^4x=N\sqrt{\qq}\;\dd^3x\,\dd t\,.
\end{equation}

The first two equations of (3.40) tell us that the components $g_{0\mu}$ of the spacetime metric are directly related to $N$ and $N^i$, so that we can replace the former by the latter. As we shall see, $g_{0\mu}$, and hence $N$ and $N^a$, are not dynamical variables: they are not determined by the field equations, and must be prescribed from outside in order to build the foliation.

\subsection{Curvature Scalar}\label{sec:curvature-scalar}

Having introduced the ADM variables, we are now in a position to rewrite the scalar curvature $\fR$ in terms of them, with the aim of rewriting the variational principle. As a first step, we relate the extrinsic curvature defined in Sec.~3.3 to the ADM variables. Since $n_a=-N\nabla_at$, we have $\nabla_cn_d=-\nabla_cN\,\nabla_dt-N\nabla_c\nabla_dt$; the second term is symmetric, and the first is proportional to $n_d$ and is annihilated by the projection on the index $d$. The extrinsic curvature is therefore one half of the projected Lie derivative of the metric along the normal:
\begin{equation}
K_{ab}=\qq_a{}^c\,\qq_b{}^d\,\nabla_{(c}n_{d)}=\tfrac12\,\qq_a{}^c\,\qq_b{}^d\,\Lie_n g_{cd}=\tfrac12\,\qq_a{}^c\,\qq_b{}^d\,\Lie_n \qq_{cd}\,,
\end{equation}
where in the last step we used $\qq_{cd}=g_{cd}+n_cn_d$. Now, with $t^a=Nn^a+N^a$ we have $\Lie_t\qq_{cd}=N\Lie_n\qq_{cd}+2\qq_{e(c}n^e\nabla_{d)}N+\Lie_{\vec N}\qq_{cd}$, where the middle term vanishes, and $\qq_a{}^c\qq_b{}^d\,\Lie_{\vec N}\qq_{cd}=\D_aN_b+\D_bN_a$. Defining $\dot \qq_{ab}\equiv\qq_a{}^c\qq_b{}^d\,\Lie_t\qq_{cd}$, whose components in adapted coordinates are $\pd_t\qq_{ij}$, we obtain
\begin{equation}\label{eq:K-ADM}
\begin{split}
\dot \qq_{ab}&=2NK_{ab}+\D_aN_b+\D_bN_a\,,\\
K_{ab}&=\frac{1}{2N}\big(\dot \qq_{ab}-\D_aN_b-\D_bN_a\big)\,.
\end{split}
\end{equation}
The same result can be obtained in components, directly from (3.35): in adapted coordinates $n_\mu=-N(1,0,0,0)$, so that $K_{ij}=\pd_in_j-\fG^\mu{}_{ij}n_\mu=N\fG^0{}_{ij}=N\big(g^{00}\fG_{0ij}+g^{0k}\fG_{kij}\big)$, and using (3.41), (3.26) and (3.25) one recovers (3.44). Note that $K_{ab}$ does not depend on the derivatives of $N$ and $N^a$ with respect to $t$.

Next, we rewrite the projection onto the hypersurface of the spacetime Riemann tensor, $\fR_{abc}{}^d$, in terms of the Riemann tensor $\sR_{abc}{}^d$ intrinsic to the hypersurface (constructed from the covariant derivative $\D_a$) and of the extrinsic curvature~\cite{SotoThesis}. Recall that the Riemann tensor is defined by
\begin{equation}
(\nabla_a\nabla_b-\nabla_b\nabla_a)\,\omega_c=\fR_{abc}{}^d\,\omega_d\,,\qquad\text{or}\qquad
(\nabla_a\nabla_b-\nabla_b\nabla_a)\,v^d=-\fR_{abc}{}^d\,v^c\,,
\end{equation}
for any covector field $\omega_c$ and vector field $v^d$, and that $\sR_{abc}{}^d$ is defined in the same way from $\D_a$ acting on spatial tensors.\footnote{The original version used $[\nabla_\alpha,\nabla_\beta]A^\mu=R^\mu{}_{\nu\alpha\beta}A^\nu$, as in~\cite{MTW}. The two conventions differ only in the order of the indices, $R^d{}_{cab}\big|_{\text{\cite{MTW}}}=R_{bac}{}^d\big|_{\text{here}}$, and give the same Ricci tensor and scalar curvature.}

To compute it we use the basis vectors $\e{i}{a}$ of adapted coordinates. Since they are coordinate vector fields they commute, $[e_i,e_j]^a=0$, and therefore $\e{i}{a}\nabla_a(\e{j}{b}\nabla_b\e{k}{c})-\e{j}{a}\nabla_a(\e{i}{b}\nabla_b\e{k}{c})=-\fR_{abd}{}^c\,\e{i}{a}\e{j}{b}\e{k}{d}$. The left-hand side can be evaluated with the Gauss formula~\eqref{eq:Gauss-formula} and with $\e{i}{a}\nabla_an^c=K_i{}^l\,\e{l}{c}$, which follows from $\qq_a{}^c\nabla_cn^b=K_a{}^b$. We obtain
\begin{equation}
\begin{split}
\e{i}{a}\nabla_a\big(\e{j}{b}\nabla_b\e{k}{c}\big)&=\e{i}{a}\nabla_a\big(\sG^l{}_{kj}\,\e{l}{c}+K_{jk}\,n^c\big)\\
&=\pd_i\sG^l{}_{kj}\,\e{l}{c}+\sG^l{}_{kj}\,\e{i}{a}\nabla_a\e{l}{c}+\pd_iK_{jk}\,n^c+K_{jk}\,\e{i}{a}\nabla_an^c\\
&=\pd_iK_{jk}\,n^c+K_{jk}K_i{}^l\,\e{l}{c}+\pd_i\sG^l{}_{kj}\,\e{l}{c}
+\sG^l{}_{kj}\big(\sG^m{}_{li}\,\e{m}{c}+K_{il}\,n^c\big)\,.
\end{split}
\end{equation}
Computing $\e{j}{a}\nabla_a(\e{i}{b}\nabla_b\e{k}{c})$ in the same way and taking the difference we have
\begin{equation}
\begin{split}
-\fR_{abd}{}^c\,\e{i}{a}\e{j}{b}\e{k}{d}
={}&n^c\big[\pd_iK_{jk}-\pd_jK_{ik}+\sG^l{}_{kj}K_{il}-\sG^l{}_{ki}K_{jl}\big]\\
&+\big[K_{jk}K_i{}^l-K_{ik}K_j{}^l+\pd_i\sG^l{}_{kj}-\pd_j\sG^l{}_{ki}+\sG^m{}_{kj}\sG^l{}_{mi}-\sG^m{}_{ki}\sG^l{}_{mj}\big]\e{l}{c}\\
={}&n^c\big(\D_iK_{jk}-\D_jK_{ik}\big)+\big(-\sR_{ijk}{}^l+K_{jk}K_i{}^l-K_{ik}K_j{}^l\big)\e{l}{c}\,,
\end{split}
\end{equation}
where in the last step we added and subtracted $\sG^l{}_{ij}K_{lk}$ (which is symmetric in $i,j$) to complete the covariant derivatives, and used the expression of the components of the intrinsic Riemann tensor in terms of the Christoffel symbols, $-\sR_{ijk}{}^l=\pd_i\sG^l{}_{kj}-\pd_j\sG^l{}_{ki}+\sG^m{}_{kj}\sG^l{}_{mi}-\sG^m{}_{ki}\sG^l{}_{mj}$, which follows from (3.45).

Equation (3.47) directly gives the projections onto the hypersurface and onto the normal. Since the $\e{i}{a}$ span $T_p\Sig$, these are equivalent to the following identities between spatial tensors:
\begin{equation}\label{eq:Gauss}
\qq_a{}^e\,\qq_b{}^f\,\qq_c{}^g\,\qq^d{}_h\,\fR_{efg}{}^h=\sR_{abc}{}^d+K_{ac}K_b{}^d-K_{bc}K_a{}^d\,,
\end{equation}
which is known as the \emph{Gauss equation} (sometimes called the Gauss--Codazzi equation), and
\begin{equation}\label{eq:Codazzi}
\qq_a{}^e\,\qq_b{}^f\,\qq_c{}^g\,\fR_{efg}{}^d\,n_d=\D_aK_{bc}-\D_bK_{ac}\,,
\end{equation}
which is the \emph{Codazzi--Mainardi equation}.

The scalar curvature is the complete contraction of the Riemann tensor. Writing $g^{ab}=\qq^{ab}-n^an^b$ we can separate it into a purely spatial part and a part with two normal components (the part with four normal components vanishes by the antisymmetry of the Riemann tensor):
\begin{equation}
\fR=\qq^{ac}\qq^{bd}\,\fR_{abcd}-2\,\fR_{ab}\,n^an^b\,,
\end{equation}
where $\fR_{abcd}=\fR_{abc}{}^e g_{ed}$. We already know the first term: contracting~\eqref{eq:Gauss} we find $\qq^{ac}\qq^{bd}\,\fR_{abcd}=\sR+K^2-K_{ab}K^{ab}$, where $K=\qq^{ab}K_{ab}$. We still have to find the second term. For that we apply the definition of the Riemann tensor to the unit normal $n^a$ of the foliation:
\begin{equation}
(\nabla_a\nabla_b-\nabla_b\nabla_a)\,n^c=-\fR_{abd}{}^c\,n^d\,,
\end{equation}
which, contracting $a$ with $c$, gives
\begin{equation}
\fR_{bd}\,n^d=\nabla_a\nabla_bn^a-\nabla_b\nabla_an^a\,.
\end{equation}
If we now contract with $n^b$ we get
\begin{equation}
\begin{split}
\fR_{ab}\,n^an^b&=n^b\nabla_a\nabla_bn^a-n^b\nabla_b\nabla_an^a\\
&=\nabla_a(n^b\nabla_bn^a)-(\nabla_an^b)(\nabla_bn^a)-\nabla_b(n^b\nabla_an^a)+(\nabla_bn^b)(\nabla_an^a)\\
&=-\nabla_a\big(n^a\nabla_bn^b-n^b\nabla_bn^a\big)+(\nabla_an^a)^2-(\nabla_an^b)(\nabla_bn^a)\,.
\end{split}
\end{equation}
To evaluate the last two terms, note that, since $n^b\nabla_an_b=0$, the decomposition~\eqref{eq:projector} gives $\nabla_an_b=K_{ab}-n_aa_b$, where $a_b\equiv n^c\nabla_cn_b$ is the acceleration of the normal, a spatial vector. Hence
\begin{equation}
\nabla_an^a=K\,,\qquad (\nabla_an^b)(\nabla_bn^a)=K_{ab}K^{ab}\,,
\end{equation}
and we can write
\begin{equation}
-\fR_{ab}\,n^an^b=K_{ab}K^{ab}-K^2+\nabla_a\Delta^a\,,
\end{equation}
with $\Delta^a\equiv n^a\nabla_bn^b-n^b\nabla_bn^a=Kn^a-a^a$. Using (3.50), (3.48) and (3.55) we can then express the scalar curvature as
\begin{equation}\label{eq:R-decomposition}
\begin{split}
\fR&=\sR+K^2-K_{ab}K^{ab}+2\big(K_{ab}K^{ab}-K^2+\nabla_a\Delta^a\big)\\
&=\sR+K_{ab}K^{ab}-K^2+2\nabla_a\Delta^a\,.
\end{split}
\end{equation}

In this section we have expressed both the Riemann tensor and the scalar curvature in terms of the new variables. With them one can rewrite the Einstein tensor $G_{ab}=R_{ab}-\tfrac12g_{ab}R$, and thus Einstein's equations in vacuum, $G_{ab}=0$. This is the procedure used to solve the equations numerically (see~\cite{Matzner}). We will not follow this path, since we are interested in constructing a Hamiltonian, for which we need a variational principle.

\subsection{Variational Principle}

Having defined the ADM variables and rewritten the scalar curvature, as well as the invariant volume element, in terms of them, we can now reformulate the variational principle that gives rise to the field equations. The action of the gravitational field is
\begin{equation}
S=\int \dd^4x\,\sqrt{-g}\;\fR\,,
\end{equation}
which, using the results of the previous sections, can be re-expressed as
\begin{equation}\label{eq:S-ADM}
S[\qq_{ab},N,N^a]=\int \dd t\int \dd^3x\,\sqrt{\qq}\,N\big(K_{ab}K^{ab}-K^2+\sR+2\nabla_a\Delta^a\big)\,.
\end{equation}

We stress that not only has the form of the Lagrangian changed in the new formulation; the action integral itself must also be reinterpreted. In (3.57) the integral is over the whole manifold, that is, over all of spacetime. When the variation is performed, the terms evaluated on the boundary are cancelled by requiring that the variation of the metric, $\delta g_{ab}$, and of its derivatives vanish on the boundary of the manifold (that is, the values of $g_{ab}$ and of its derivatives are fixed there). In general one does not know a priori the topology of the manifold $M$, so it is not known beforehand whether it has a boundary or not.

Expression (3.58) has a different interpretation, since it contains two integrations of a different nature. The first is over the hypersurface $\Sig$, so that the Lagrangian is given by an integral over the hypersurface,
\begin{equation}
L=\int \dd^3x\,\sqrt{\qq}\,N\big(K_{ab}K^{ab}-K^2+\sR+2\nabla_a\Delta^a\big)\,,
\end{equation}
and the action is the integral of the Lagrangian between an initial value $t_0$ and a final value $t_1$ of the parameter:
\begin{equation}
S=\int_{t_0}^{t_1}\dd t\,L\,.
\end{equation}
In other words, one specifies the metric on the initial ($t_0$) and on the final ($t_1$) hypersurface, and the variation is over all possible hypersurfaces (their metrics) connecting the two. The divergence term splits into two parts. Since $\sqrt{-g}\,\nabla_a\Delta^a=\pd_\mu(\sqrt{-g}\,\Delta^\mu)$, and in adapted coordinates $\sqrt{-g}\,\Delta^0=\sqrt{\qq}\,K$, it gives a total time derivative, $2\,\frac{\dd}{\dd t}\int\dd^3x\,\sqrt{\qq}\,K$, which contributes only on the initial and final hypersurfaces, and a total spatial divergence, $2\,\pd_i(\sqrt{-g}\,\Delta^i)$, which becomes an integral over the boundary of $\Sig$. These terms do not contribute to the local field equations, and we shall drop them.\footnote{Some care is needed here. Since $K$ contains $\dot\qq_{ab}$, fixing $\qq_{ab}$ on the initial and final hypersurfaces does not by itself make the variation of the total time derivative vanish; this term is precisely cancelled by the boundary term of York~\cite{York} and Gibbons and Hawking~\cite{GibbonsHawking}, which must be added to (3.57) to obtain a well-posed variational principle in which only the induced metric is fixed on the boundary. When $\Sig$ is not compact, the surface terms on $\pd\Sig$ require a separate treatment. Both issues are discussed in Sec.~\ref{sec:ADM-energy}.}

It is worth stressing that this expression of the gravitational action restricts the topology of spacetime, which must be of the form $\Sig\times\mathbb{R}$, and this topology is maintained throughout the evolution of the hypersurfaces. That is, changes in the topology of the hypersurfaces are not allowed (for example, a universe that is closed will always be closed, and one that is open will always be open).

\subsection{Hamiltonian Formulation}\label{sec:hamiltonian}

So far we have rewritten the action functional in such a way that the field equations in vacuum follow from requiring the action to be stationary, $\delta S=0$. The Lagrangian density contains second spatial derivatives of $\qq_{ab}$, so in principle the Euler--Lagrange equations could be of fourth order. However, these second derivatives appear linearly and, when the variation is performed (see~\cite{MTW} and~\cite{Landau}), the terms with higher derivatives combine into boundary terms; the resulting field equations are of second order, as one expects for dynamical equations.

One can reduce the order of the field equations, so as to have only first-order equations, in several ways. The first consists in taking the Palatini action~\cite{MTW}, which treats both the ten components of $g_{ab}$ and the forty components of a symmetric affine connection $\Gamma^c{}_{ab}$ as independent variables and varies with respect to all of them. With this procedure one obtains the same Einstein equations, as well as the relation between the metric and the connection.

The other route consists in defining new variables, the momenta, starting from the Lagrangian, and rewriting the Lagrangian in terms of the original variables and the momenta, taking both as independent variables (so that the number of independent variables is doubled). One of the intermediate steps is the construction of the Hamiltonian. This last step is in general not straightforward when there are constraints among the variables. To deal with such systems a whole theory has been developed, that of singular, or constrained, systems (see~\cite{DiracLectures}, \cite{Sundermeyer} and~\cite{HRT}).

With the formalism developed so far, the Hamiltonian can be constructed directly since, as explained before, the Lagrangian density does not depend on the time derivatives of $N$ and $N^a$, which are therefore non-dynamical. We define the momenta canonically conjugate to the six components of $\qq_{ab}$ by
\begin{equation}
\pi^{ab}=\frac{\pd\mathcal{L}}{\pd\dot \qq_{ab}}\,,
\end{equation}
while the momenta associated with $N$ and $N^a$ clearly vanish,
\begin{equation}
P\equiv\frac{\pd\mathcal{L}}{\pd\dot N}=0\,,\qquad P_a\equiv\frac{\pd\mathcal{L}}{\pd\dot N^a}=0\,.
\end{equation}
In order to compute the momenta (3.61) we introduce an object with four indices,
\begin{equation}
G^{abcd}\equiv\sqrt{\qq}\,\Big(\tfrac12\big(\qq^{ac}\qq^{bd}+\qq^{ad}\qq^{bc}\big)-\qq^{ab}\qq^{cd}\Big)\,,
\end{equation}
called the \emph{supermetric}. We can then write the Lagrangian density in the form
\begin{equation}
\mathcal{L}=N\big(G^{abcd}K_{ab}K_{cd}+\sqrt{\qq}\;\sR\big)\,.
\end{equation}
From the definition of the momenta and the previous expression,
\[
\pi^{ab}=\frac{\pd\mathcal{L}}{\pd\dot \qq_{ab}}=2N\,G^{efcd}K_{cd}\,\frac{\pd K_{ef}}{\pd\dot \qq_{ab}}\,,
\]
and from Eq.~(3.44) it follows that
\begin{equation}
\frac{\pd K_{ef}}{\pd\dot \qq_{ab}}=\frac{1}{2N}\cdot\frac12\big(\delta^a{}_e\delta^b{}_f+\delta^a{}_f\delta^b{}_e\big)\equiv\frac{1}{2N}\,\delta^{ab}{}_{ef}\,,
\end{equation}
so that
\begin{equation}\label{eq:pi-K}
\pi^{ab}=G^{abcd}K_{cd}=\sqrt{\qq}\,\big(K^{ab}-K\qq^{ab}\big)\,.
\end{equation}
As can be seen from the definition of the supermetric and from the last expression, the momenta $\pi^{ab}$ are tensor densities of weight one on the hypersurface.

To perform the Legendre transformation of the Lagrangian and define the Hamiltonian, we must express the velocities as functions of the momenta. The velocities $\dot N$ and $\dot N^a$ are arbitrary functions and therefore cannot be expressed in terms of the coordinates and momenta. The velocities $\dot \qq_{ab}$, on the other hand, can, since Eq.~(3.66) can be inverted to express $K_{ab}$ in terms of the momenta. For that we need the inverse of the supermetric, that is, the object $G_{abcd}$ which, contracted with the supermetric, gives the identity on symmetric tensors,
\begin{equation}
G_{abcd}\,G^{cdef}=\delta_{ab}{}^{ef}\,.
\end{equation}
The most general form of the inverse is
\begin{equation}
G_{abcd}=\frac{1}{\sqrt{\qq}}\Big(\frac{A}{2}\big(\qq_{ac}\qq_{bd}+\qq_{ad}\qq_{bc}\big)+B\,\qq_{ab}\qq_{cd}\Big)\,.
\end{equation}
Inserting (3.68) into (3.67) one finds $A=1$, $B=-\tfrac12$. The extrinsic curvature can now be expressed as
\begin{equation}
K_{ab}=G_{abcd}\,\pi^{cd}\,.
\end{equation}
From Eqs.~(3.44) and (3.69) we can write the velocities $\dot \qq_{ab}$ in the form
\begin{equation}\label{eq:qdot}
\begin{split}
\dot \qq_{ab}&=\D_aN_b+\D_bN_a+2NK_{ab}\\
&=\D_aN_b+\D_bN_a+2N\,G_{abcd}\,\pi^{cd}\,.
\end{split}
\end{equation}

We are now in a position to write the Hamiltonian, where the Legendre transformation is performed only on the velocities $\dot \qq_{ab}$:
\begin{equation}
H=\int \dd^3x\,\big(\pi^{ab}\dot \qq_{ab}-\mathcal{L}\big)\,.
\end{equation}
The Hamiltonian density can now be expressed as
\begin{equation}
\begin{split}
\pi^{ab}\dot \qq_{ab}-\mathcal{L}&=G^{abcd}K_{cd}\big(\D_aN_b+\D_bN_a+2NK_{ab}\big)-N\big(G^{abcd}K_{ab}K_{cd}+\sqrt{\qq}\;\sR\big)\\
&=N\sqrt{\qq}\,\big(K_{ab}K^{ab}-K^2-\sR\big)+2\pi^{ab}\D_aN_b\,.
\end{split}
\end{equation}
The second term can be integrated by parts (taking into account that $\pi^{ab}$ is a tensor density of weight one, so that $\D_a(\pi^{ab}N_b)=\pd_a(\pi^{ab}N_b)$), after which, discarding a surface term,\footnote{For compact $\Sig$ this term is absent. When $\Sig$ is not compact, as for asymptotically flat spacetimes, the surface terms do not vanish in general, and they must be treated as shown by Regge and Teitelboim~\cite{ReggeTeitelboim}: they give rise to the energy and momentum of the spacetime (Sec.~\ref{sec:ADM-energy}).} we are left with the Hamiltonian
\begin{equation}\label{eq:H}
\begin{split}
H&=\int \dd^3x\,\Big[N\sqrt{\qq}\,\big(K_{ab}K^{ab}-K^2-\sR\big)-2N_b\,\D_a\pi^{ab}\Big]\\
&\equiv\int \dd^3x\,\big(N\Hs+N^a\Hv_a\big)\,,
\end{split}
\end{equation}
where
\begin{align*}
\Hs&=\sqrt{\qq}\,\big(K_{ab}K^{ab}-K^2-\sR\big)=G_{abcd}\,\pi^{ab}\pi^{cd}-\sqrt{\qq}\;\sR
=\frac{1}{\sqrt{\qq}}\Big(\pi_{ab}\pi^{ab}-\tfrac12\pi^2\Big)-\sqrt{\qq}\;\sR\,,\\
\Hv_a&=-2\,\qq_{ac}\,\D_b\pi^{bc}\,,
\end{align*}
with $\pi=\qq_{ab}\pi^{ab}$; in the first expression for $\Hs$, $K_{ab}$ stands for $G_{abcd}\pi^{cd}$.

We can then write the first-order Lagrangian density, from (3.71) and (3.73), as
\begin{equation}\label{eq:L-first-order}
\mathcal{L}(\qq_{ab},\dot \qq_{ab},\pi^{ab},N,N^a)=\pi^{ab}\dot \qq_{ab}-N\Hs-N^a\Hv_a\,.
\end{equation}
The action integral obtained from this Lagrangian density is a functional of the spatial metric $\qq_{ab}$, of the momenta $\pi^{ab}$ and of the functions $N$ and $N^a$. Two points should be noted about the form of (3.74): the functions $N$ and $N^a$ play the role of Lagrange multipliers, and the action is written in parametrized form. Let us clarify these points. If we vary the action with respect to $N$ and $N^a$ we find $\Hs=0$ and $\Hv_a=0$. Thus $N$ and $N^a$ multiply quantities that must vanish: they are Lagrange multipliers and, in principle, arbitrary functions, a reflection of the freedom in the choice of the foliation and of the coordinates on each hypersurface. The equations of motion for the other variables are obtained by varying the action with respect to them; for instance, variation with respect to $\pi^{ab}$ gives back (3.70), as it should. The variables $\qq_{ab}$ and $\pi^{ab}$ are not independent of each other, since they are related by the four conditions $\Hs=0$, $\Hv_a=0$, the constraints of the theory. The theory is in parametrized form because the first-order Lagrangian is not in canonical form, which is the form it would have if its Hamiltonian were a function of the canonical variables $(\qq_{ab},\pi^{ab})$ alone. Here the Hamiltonian also contains $N$ and $N^a$, and it vanishes when the constraints are satisfied; this is what happens when a theory is written in parametrized form.

To illustrate the distinction between the canonical and the parametrized form, let us consider a non-relativistic particle with generalized coordinates $q^i$ and velocities $\dd q^i/\dd t$. Its first-order action is
\begin{equation}
S=\int_{t_0}^{t_1}\dd t\,L=\int_{t_0}^{t_1}\dd t\,\Big(\frac{\dd q^i}{\dd t}\,p_i-H(q^i,p_i)\Big)\,,
\end{equation}
which is in canonical form. If we now take time as an additional coordinate $q^0$, and regard all the coordinates as functions of a parameter $\tau$, we can identify the Hamiltonian with (minus) the momentum canonically conjugate to the coordinate $q^0$:
\begin{equation}
p_0=-H(q^i,p_i)\,.
\end{equation}
It follows that the momentum $p_0$ is not independent of the other variables: Eq.~(3.76) is a constraint among them. Once time has lost its special status, the action (3.75) can be rewritten as
\begin{equation}
S=\int_{\tau_0}^{\tau_1}\dd\tau\;\dot q^\alpha p_\alpha\,,
\end{equation}
where the dot now denotes $\dd/\dd\tau$ and $\alpha=0,1,\ldots,n$. The variables, however, are subject to the constraint (3.76), which can be written as $\mathcal{C}\equiv p_0+H(q^i,p_i)=0$. This constraint must be added to the Lagrangian, multiplied by an arbitrary multiplier $N$, to obtain the complete Lagrangian. Let us therefore consider
\begin{equation}
L=\dot q^\alpha p_\alpha-N\mathcal{C}\,.
\end{equation}
This Lagrangian is in parametrized form. To bring it to canonical form one has to retrace these steps. The first step is to eliminate $\mathcal{C}$, i.e.\ to solve the constraint by setting $p_0=-H$. The action then becomes
\begin{equation}
S=\int_{\tau_0}^{\tau_1}\dd\tau\,\big(\dot q^i p_i-H\,\dot q^0\big)\,,
\end{equation}
which can be rewritten in such a way that the dependence on the parameter $\tau$ disappears:
\begin{equation}
S=\int \dd q^0\,\Big(\frac{\dd q^i}{\dd q^0}\,p_i-H\Big)\,.
\end{equation}
This form of the action does not depend on the parameter $\tau$ (which no longer appears in it) and is therefore unchanged by a change of parameter.

In practice, going from (3.79) to (3.80) amounts to specifying a relation between the parameter and the coordinate $q^0$. This is possible because of the freedom to choose the parameter arbitrarily (a gauge freedom). If we impose a coordinate condition such as $q^0=\tau$, the action takes the form (3.80) with the change of notation $q^0\to\tau$. This example shows that, in general terms, the way to reduce a parametrized action to canonical form is to insert the solution of the constraints and to impose coordinate conditions.

The gravitational action has the form (3.78). To solve the equations in this way, the first step would be to transform it to canonical form, the analogue of (3.75); in other words, one would have to solve the constraints and impose coordinate conditions. This has been possible only in very particular cases (see the applications), and in general it has been achieved only formally~\cite{Sundermeyer}.

\subsection{Einstein's Equations in 3+1 Form}\label{sec:einstein-3+1}

\emph{Sections~3.8--3.10 are new in the present edition.} In the introduction we stated that some of Einstein's equations are constraints on the dynamical variables, while the rest describe their evolution. With the results of Secs.~\ref{sec:curvature-scalar} and~\ref{sec:hamiltonian} we can now make this statement precise.

Consider first the components of the Einstein tensor along the normal. Since $g_{ab}n^an^b=-1$, we have $G_{ab}n^an^b=\fR_{ab}n^an^b+\tfrac12\fR$, and comparing with (3.50) and with the contraction of the Gauss equation (3.48) we find
\begin{equation}\label{eq:Ham-constraint-G}
2\,G_{ab}\,n^an^b=\sR+K^2-K_{ab}K^{ab}\,.
\end{equation}
Similarly, contracting the Codazzi--Mainardi equation (3.49) with $\qq^{ac}$, and using that the contraction of $\fR_{abcd}$ with three normals vanishes, we obtain
\begin{equation}\label{eq:mom-constraint-G}
G_{ab}\,n^a\,\qq^b{}_c=\D_bK^b{}_c-\D_cK\,.
\end{equation}
Neither expression contains time derivatives of $K_{ab}$, that is, second time derivatives of the metric: these four equations restrict the data $(\qq_{ab},K_{ab})$ on each hypersurface. Comparing with (3.66) and (3.73), they are precisely the constraints of the Hamiltonian formulation,
\begin{equation}\label{eq:constraints-G}
\Hs=-2\sqrt{\qq}\;G_{ab}\,n^an^b\,,\qquad \Hv_a=-2\sqrt{\qq}\;\qq_a{}^b\,G_{bc}\,n^c\,,
\end{equation}
so that the four equations $G_{ab}n^b=0$ are equivalent to $\Hs=0$ and $\Hv_a=0$. In the presence of matter, with $G_{ab}=\kappa T_{ab}$, they read $\sR+K^2-K_{ab}K^{ab}=2\kappa\,T_{ab}n^an^b$ and $\D_bK^b{}_c-\D_cK=\kappa\,\qq_c{}^bT_{ba}n^a$. The contracted Bianchi identity (1.2) is what guarantees that these four equations are preserved by the evolution generated by the remaining six, as anticipated in the introduction.

The remaining six equations are evolution equations. In the Hamiltonian formulation they are Hamilton's equations,
\begin{equation}\label{eq:Hamilton-eqs}
\dot\qq_{ab}=\frac{\delta H}{\delta\pi^{ab}}\,,\qquad \dot\pi^{ab}=-\frac{\delta H}{\delta \qq_{ab}}\,,
\end{equation}
with $H$ given by (3.73). The first reproduces (3.70). The second requires the variation of the scalar curvature, $\delta\,\sR=-\sR^{ab}\delta \qq_{ab}+\D^a\D^b\delta \qq_{ab}-\D^c\D_c(\qq^{ab}\delta \qq_{ab})$, and a lengthy but straightforward computation~\cite{ADM,Wald} gives
\begin{equation}\label{eq:pidot}
\begin{split}
\dot\pi^{ab}={}&-N\sqrt{\qq}\,\big(\sR^{ab}-\tfrac12\sR\,\qq^{ab}\big)
+\frac{N}{2\sqrt{\qq}}\,\qq^{ab}\big(\pi_{cd}\pi^{cd}-\tfrac12\pi^2\big)
-\frac{2N}{\sqrt{\qq}}\big(\pi^{ac}\pi_c{}^b-\tfrac12\pi\,\pi^{ab}\big)\\
&+\sqrt{\qq}\,\big(\D^a\D^bN-\qq^{ab}\D_c\D^cN\big)+\D_c\big(\pi^{ab}N^c\big)-2\pi^{c(a}\D_cN^{b)}\,,
\end{split}
\end{equation}
where $\dot\pi^{ab}$ is the time derivative of the components of $\pi^{ab}$ in adapted coordinates; the last two terms are the Lie derivative of the density $\pi^{ab}$ along $N^a$. To see how (3.85) is related to Einstein's equations, recall that the action built from (3.74) differs from the Einstein--Hilbert action only by boundary terms, and that varying $\qq_{ab}$ at fixed $N$ and $N^a$ amounts to varying $g_{ab}$ by $\delta g_{ab}=\qq_a{}^c\qq_b{}^d\,\delta \qq_{cd}$. Since $\delta S=-\int \dd^4x\sqrt{-g}\,G^{ab}\delta g_{ab}$, one finds that, for an arbitrary metric,
\begin{equation}\label{eq:pidot-G}
\dot\pi^{ab}+\frac{\delta H}{\delta \qq_{ab}}=N\sqrt{\qq}\;\qq^a{}_c\,\qq^b{}_d\,G^{cd}\,.
\end{equation}
Equation (3.85) therefore holds if and only if the spatial projection of Einstein's equations holds. Together with (3.83), this shows that the ten Einstein equations are equivalent to the four constraints together with the six Hamiltonian evolution equations, which is the precise content of the statement made in the introduction.

In numerical relativity one works instead with an evolution equation for the extrinsic curvature, in a form due to York~\cite{York79}. Defining $\dot K_{ab}\equiv \qq_a{}^c\qq_b{}^d\,\Lie_tK_{cd}$, a computation analogous to that of Sec.~\ref{sec:curvature-scalar} gives, for an arbitrary metric,
\begin{equation}\label{eq:Kdot}
\dot K_{ab}=\Lie_{\vec N}K_{ab}+\D_a\D_bN-N\big(\sR_{ab}+K\,K_{ab}-2K_{ac}K^c{}_b\big)+N\,\qq_a{}^c\qq_b{}^d\,\fR_{cd}\,.
\end{equation}
Equations (3.70) and (3.87) with $\fR_{ab}=0$, together with the constraints (3.81)--(3.82), are what are usually called the ``ADM equations'', although in this form they are due to York. Note that (3.87) follows from the spatial projection of $\fR_{ab}=0$ rather than of $G_{ab}=0$. A short computation using $\fR=G_{ab}n^an^b-\qq^{ab}G_{ab}$ shows that $\qq_a{}^c\qq_b{}^d\fR_{cd}=0$ is equivalent to $\qq_a{}^c\qq_b{}^dG_{cd}=\qq_{ab}\,G_{cd}n^cn^d$: York's evolution system is that of ADM with a multiple of $\qq_{ab}$ times the Hamiltonian constraint added. The two systems agree on solutions of the constraints, but not away from them, a difference that matters in numerical work (Sec.~\ref{sec:NR}).

\subsection{Constraint Analysis and Degrees of Freedom}\label{sec:constraint-algebra}

In Sec.~\ref{sec:hamiltonian} we mentioned that systems with constraints require a theory of their own, due to Dirac and Bergmann~\cite{DiracLectures,Sundermeyer,HRT,HenneauxTeitelboim}. We can now apply it to general relativity.

The canonical variables are $\qq_{ab}$ and $\pi^{ab}$, together with $N$, $N^a$ and their momenta $P$, $P_a$, with Poisson brackets
\begin{equation}
\begin{gathered}
\{\qq_{ab}(x),\pi^{cd}(y)\}=\delta_{(a}{}^c\,\delta_{b)}{}^d\,\delta^3(x,y)\,,\\
\{N(x),P(y)\}=\delta^3(x,y)\,,\qquad
\{N^a(x),P_b(y)\}=\delta^a{}_b\,\delta^3(x,y)\,,
\end{gathered}
\end{equation}
all others vanishing. The primary constraints $P\approx0$, $P_a\approx0$ (the symbol $\approx$ denotes an equality that holds on the constraint surface) are added to the Hamiltonian with arbitrary multipliers $\lambda$, $\lambda^a$, giving the total Hamiltonian
\begin{equation}
H_T=\int \dd^3x\,\big(N\Hs+N^a\Hv_a+\lambda P+\lambda^aP_a\big)\,.
\end{equation}
Consistency requires the primary constraints to be preserved in time:
\begin{equation}
\dot P=\{P,H_T\}=-\Hs\approx0\,,\qquad \dot P_a=\{P_a,H_T\}=-\Hv_a\approx0\,.
\end{equation}
These are the secondary constraints: the Hamiltonian and momentum constraints, which we found before by varying $N$ and $N^a$. To find out whether new constraints arise we must compute the Poisson brackets of $\Hs$ and $\Hv_a$ among themselves. It is convenient to smear them with test functions,
\begin{equation}\label{eq:smeared}
H[N]\equiv\int \dd^3x\,N\Hs\,,\qquad D[\vec N]\equiv\int \dd^3x\,N^a\Hv_a=\int \dd^3x\;\pi^{ab}\,\Lie_{\vec N}\qq_{ab}\,,
\end{equation}
where the last equality follows from (3.73) after an integration by parts (we discard surface terms here; they are the subject of Sec.~\ref{sec:ADM-energy}). With the Poisson bracket of two functionals,
$\{F,G\}=\int \dd^3x\,\big(\frac{\delta F}{\delta \qq_{ab}}\frac{\delta G}{\delta\pi^{ab}}-\frac{\delta F}{\delta\pi^{ab}}\frac{\delta G}{\delta \qq_{ab}}\big)$, the second expression in (3.91) shows that $D[\vec N]$ generates spatial diffeomorphisms:
\begin{equation}\label{eq:D-generates}
\{\qq_{ab},D[\vec N]\}=\Lie_{\vec N}\qq_{ab}\,,\qquad \{\pi^{ab},D[\vec N]\}=\Lie_{\vec N}\pi^{ab}\,.
\end{equation}
Using (3.92) and the variation of $\sR$ quoted above one finds~\cite{Dirac58,DeWitt,Teitelboim73,HKT}
\begin{equation}\label{eq:algebra}
\begin{split}
\{D[\vec N],D[\vec M]\}&=D\big[[\vec N,\vec M]\big]\,,\\
\{D[\vec N],H[M]\}&=H\big[\Lie_{\vec N}M\big]\,,\\
\{H[N],H[M]\}&=D[\vec V]\,,\qquad V^a=\qq^{ab}\big(N\pd_bM-M\pd_bN\big)\,.
\end{split}
\end{equation}
The first two brackets express that $\Hv_a$ and $\Hs$ transform as densities under spatial diffeomorphisms. In the third, only the terms of $\delta H[N]/\delta \qq_{ab}$ with derivatives of $N$ survive the antisymmetrization, and an integration by parts produces the momentum constraint (Exercise~A.7). The brackets close: they vanish on the constraint surface, so no new constraints arise, and the multipliers $N$, $N^a$, $\lambda$, $\lambda^a$ remain arbitrary. All the constraints are therefore first class. Note, however, that (3.93) is not a Lie algebra, since the last bracket involves the metric $\qq^{ab}$: its ``structure constants'' are functions on phase space. We return to the meaning of this algebra in Sec.~\ref{sec:HKT}.

First-class constraints generate gauge transformations. $P$ and $P_a$ generate arbitrary changes of $N$ and $N^a$, which reflects the fact that the lapse and the shift are not dynamical. $H[N]$ and $D[\vec N]$ generate, on solutions of the field equations, the change of the canonical data under a deformation of the hypersurface by the spacetime vector field $Nn^a+N^a$, normal and tangential respectively. This is the canonical expression of the general covariance of the theory.

We can now count the degrees of freedom. At each point of $\Sig$ the pair $(\qq_{ab},\pi^{ab})$ has twelve components. Each of the four first-class constraints $\Hs$, $\Hv_a$ removes two of them, one through the constraint itself and one through the gauge freedom it generates. We are left with
\begin{equation}\label{eq:dof}
2\times6-2\times4=4=2\times2
\end{equation}
phase-space dimensions per point, that is, two configuration degrees of freedom per point of space, as anticipated in Sec.~4.1 on the basis of the linearized theory; they correspond to the two polarizations of gravitational waves. The pairs $(N,P)$ and $(N^a,P_a)$ do not contribute, being pure gauge.

\subsection{Boundary Terms, Asymptotic Flatness and the ADM Energy}\label{sec:ADM-energy}

In Secs.~3.6 and~3.7 we discarded several surface terms. This is harmless for the local field equations, but the surface terms carry important information.

The first concerns the initial and final hypersurfaces. As noted in Sec.~3.6, the divergence in (3.58) gives a total time derivative, $2\frac{\dd}{\dd t}\int \dd^3x\sqrt{\qq}\,K$, whose variation does not vanish when only $\qq_{ab}$ is held fixed at $t_0$ and $t_1$, since $K$ contains $\dot \qq_{ab}$. The remedy, due to York~\cite{York} and to Gibbons and Hawking~\cite{GibbonsHawking}, is to add a boundary term to the Einstein--Hilbert action~\cite{Poisson}:
\begin{equation}\label{eq:GHY}
S=\int_M \dd^4x\,\sqrt{-g}\;\fR+2\oint_{\pd M}\dd^3y\;\varepsilon\sqrt{|h|}\;\mathcal{K}\,,
\end{equation}
where $h$ is the determinant of the metric induced on the boundary $\pd M$, $\mathcal{K}$ is the trace of its extrinsic curvature, defined as in (3.29) with the outward unit normal, and $\varepsilon=\pm1$ is the norm of that normal. On $\Sig_{t_1}$ the outward normal is $n^a$, with $\mathcal{K}=K$, while on $\Sig_{t_0}$ it is $-n^a$, with $\mathcal{K}=-K$; in both cases $\varepsilon=-1$. The boundary term therefore equals $-2\big[\int \dd^3x\sqrt{\qq}\,K\big]_{t_0}^{t_1}$, which cancels exactly the total time derivative coming from the divergence. The variational principle is then well posed with $\qq_{ab}$ fixed on the initial and final hypersurfaces, and the action is (3.58) without the divergence term, plus contributions from the timelike part of the boundary, if there is one.

The second, more interesting, case is that of a non-compact $\Sig$, such as a hypersurface in the spacetime of an isolated system. We say that the data $(\qq_{ab},\pi^{ab})$ on $\Sig$ are asymptotically flat if, outside a compact set, there are coordinates $x^i$ in which
\begin{equation}\label{eq:falloff}
\qq_{ij}=\delta_{ij}+O(1/r)\,,\qquad \pd_k\qq_{ij}=O(1/r^2)\,,\qquad \pi^{ij}=O(1/r^2)\,,\qquad r^2=\delta_{ij}x^ix^j\,,
\end{equation}
supplemented by suitable parity conditions on the leading terms~\cite{ReggeTeitelboim,BeigOMurchadha}. Consider the Hamiltonian $H[N]+D[\vec N]$ with a lapse that tends to one and a shift that tends to zero at infinity, corresponding to a unit time translation. For $H$ to generate Hamilton's equations its variation must be of the form $\int \dd^3x\,(A^{ab}\delta \qq_{ab}+B_{ab}\delta\pi^{ab})$, with no surface terms. Using the variation of $\sR$ quoted in Sec.~\ref{sec:einstein-3+1}, however, one finds
\begin{equation}\label{eq:deltaH-surface}
\delta\big(H[N]+D[\vec N]\big)=\int \dd^3x\,\big(A^{ab}\delta \qq_{ab}+B_{ab}\delta\pi^{ab}\big)-\oint_{S_\infty}\dd S^i\,\big(\pd_j\delta \qq_{ij}-\pd_i\delta \qq_{jj}\big)\,,
\end{equation}
where $S_\infty$ is a sphere at infinity and $\dd S^i$ its outward area element in the asymptotic coordinates; all other surface terms vanish under (3.96). As shown by Regge and Teitelboim~\cite{ReggeTeitelboim}, the remedy is to add to the Hamiltonian the surface integral whose variation cancels this term,
\begin{equation}\label{eq:H-improved}
H=H[N]+D[\vec N]+E\,,\qquad E\equiv\oint_{S_\infty}\dd S^i\,\big(\pd_j\qq_{ij}-\pd_i\qq_{jj}\big)\,,
\end{equation}
which is then a differentiable function on phase space. Since the constraints vanish on solutions, the value of the Hamiltonian is given entirely by the surface term: it is the ADM energy. In ordinary units, where the action carries the factor $1/16\pi G$, the ADM energy and momentum are~\cite{ADM}
\begin{equation}\label{eq:ADM-EP}
E_{\rm ADM}=\frac{1}{16\pi G}\oint_{S_\infty}\dd S^i\,\big(\pd_j\qq_{ij}-\pd_i\qq_{jj}\big)\,,\qquad
P_i^{\rm ADM}=-\frac{1}{8\pi G}\oint_{S_\infty}\dd S^j\,\pi_{ij}\,,
\end{equation}
where $\pi_{ij}$ is the momentum (3.66) with its indices lowered with $\delta_{ij}$. The momentum is the value of the generator of asymptotic spatial translations, $D[\vec N]$ with $N^a$ tending to a constant vector, made differentiable in the same way. The minus sign arises because, by (3.92), the generator $D[\vec\xi\,]$ moves the fields by $\Lie_{\vec\xi}$, that is, it translates the field configuration by $-\xi^a$. Asymptotic rotations and boosts give the angular momentum and the centre-of-mass charge, for which the parity conditions are needed~\cite{ReggeTeitelboim,BeigOMurchadha}. The positivity of the ADM energy for matter satisfying the dominant energy condition, long conjectured, was proved by Schoen and Yau~\cite{SchoenYau} and by Witten~\cite{Witten81}.

As an example, consider data with $K_{ab}=0$ (time-symmetric data) and a conformally flat metric, $\qq_{ab}=\psi^4\delta_{ab}$. The momentum constraint is satisfied identically and, since for such a metric $\sR=-8\psi^{-5}\delta^{ij}\pd_i\pd_j\psi$, the Hamiltonian constraint reduces to Laplace's equation,
\begin{equation}\label{eq:Laplace-psi}
\delta^{ij}\pd_i\pd_j\psi=0\,.
\end{equation}
The solution $\psi=1+m/2r$ gives the hypersurfaces $t=\text{const}$ of the Schwarzschild solution in isotropic coordinates,
$\dd s^2=-\big(\frac{1-m/2r}{1+m/2r}\big)^2\dd t^2+\big(1+\frac{m}{2r}\big)^4\big(\dd r^2+r^2\dd\Omega^2\big)$, with $m=GM$. Then $\pd_j\qq_{ij}-\pd_i\qq_{jj}=-2\pd_i\psi^4=4m\,x^i/r^3+O(1/r^3)$, and (3.99) gives $E_{\rm ADM}=16\pi m/16\pi G=M$, as it should. More generally, $\psi=1+\sum_\alpha m_\alpha/(2|x-x_\alpha|)$ describes several black holes momentarily at rest (Brill--Lindquist data~\cite{BrillLindquist}), with $E_{\rm ADM}=\sum_\alpha m_\alpha/G$. Data with non-vanishing momentum and spin are obtained with the extrinsic curvature of Bowen and York~\cite{BowenYork}; they are the starting point of most simulations of binary black holes (Sec.~\ref{sec:NR}).

\section{Applications}\label{sec:applications}

In this section we briefly describe some of the fields that have been developed using the ADM formalism, as they stood in 1991; later developments are described in Sec.~\ref{sec:developments}.

\subsection{Quantum Gravity}

Until recently, one of the main motivations for building a Hamiltonian formulation (that is, for identifying canonically conjugate variables starting from a Lagrangian, and obtaining the Hamiltonian as its Legendre transform with respect to the velocities) was the possibility of quantizing the theory: associating with each canonical variable an operator on a vector space, and imposing commutation relations between pairs of conjugate variables. The Hamiltonian operator then plays the role of the generator of time translations, and it governs the dynamics of the system.

For general relativity, however, quantization does not follow directly from the Hamiltonian we have constructed, since there are four constraints~\eqref{eq:H} among the canonical variables. To overcome this difficulty one can follow two different paths, which are roughly:
\begin{enumerate}
\item[1)] To try to isolate the true degrees of freedom of the gravitational field (which, from the linearized equations, are known to be two per point of space); this is the idea followed in the ADM formalism. Once the degrees of freedom have been isolated, one proceeds to quantize. As mentioned above, this has been accomplished only formally.
\item[2)] To keep all the variables, but to impose the four constraints, now as operators, as conditions on the state vectors: the physical states are those annihilated by the constraints. This is the approach of Dirac and DeWitt.
\end{enumerate}

In recent years there have also been attempts to quantize the gravitational field using the idea of a \emph{sum over histories}: all possible evolutions of the three-geometries between two fixed three-geometries are considered, each with a weight determined by the action evaluated along it~\cite{GuvenRyan}.

Likewise, canonical formulations of general relativity alternative to that of ADM have been constructed, such as the Ashtekar formalism~\cite{Ashtekar}.

\subsection{Geometrodynamics}

Geometrodynamics is the name given to general relativity when it is regarded as a dynamical theory~\cite{Wheeler}. As mentioned in the introduction, the covariant formulation of general relativity is not the most appropriate one for appreciating its dynamical character. The introduction of the new variables, and the view of spacetime as formed by the evolution of the metric $\qq_{ab}(t)$, suggest which is the dynamical variable that evolves. However, the action (3.58) is invariant under diffeomorphisms of the hypersurface $\Sig_t$, since it does not depend on the coordinate system used on it. This suggests that the truly relevant property of each hypersurface $\Sig_t$ is not the metric defined on it, but the intrinsic geometry of the manifold itself, i.e.\ the metric up to diffeomorphisms.

This conclusion is fairly clear if one considers the example of the spheres generating Euclidean space given in Sec.~\ref{sec:3+1}. If one pictures $\mathbb{R}^3\setminus\{0\}$ as formed by infinitely many spheres glued together, it does not matter which intrinsic coordinate system is given to each sphere: it always remains a sphere.

Spacetime is, then, the evolution (or the collection, if one prefers to see it that way) of three-geometries. The object that evolves is the geometry of a three-manifold $\Sig$, and this is the dynamical variable of general relativity; hence the name geometrodynamics.

If it is the three-geometries that evolve, it is natural to ask: in what space do they evolve? The answer is almost trivial: in the space of three-geometries. This is the set whose points are all the possible three-geometries on the manifold $\Sig$, that is, the Riemannian metrics on $\Sig$ modulo diffeomorphisms. It is known as \emph{superspace}~\cite{Wheeler}, and we denote it by $\mathcal{S}$.

There is an analogy between the evolution of a particle and that of a geometry. The particle lives in spacetime and its evolution is a trajectory in it. In the same way, a three-geometry lives in superspace and its evolution is a trajectory in it. The analogy is not completely literal, however, since the ``trajectories'' in superspace that define a spacetime are not one-dimensional curves (parametrized by a real number) but submanifolds of superspace. Let us briefly see why.

Suppose we are given a spacetime. The foliations that can be built in it are infinite in number: there is one for each function $t$ on spacetime whose level sets $t=\text{const}$ are non-intersecting spacelike hypersurfaces, and clearly there are uncountably many such functions. When one of them is chosen, i.e.\ when a particular foliation is fixed, spacetime is described by a curve of hypersurfaces parametrized by the value of $t$, which is mapped to a curve in superspace. It is also clear from this argument that the set of all spacelike hypersurfaces of the given spacetime, known as \emph{hyperspace}~\cite{Kuchar}, is mapped by the induced three-geometry onto a submanifold of superspace that contains all those curves; it is this submanifold that represents the spacetime.

\subsection{Cosmology}

In the late 1960s and early 1970s the study of cosmological models, in particular of homogeneous ones, was developed with the help of the Hamiltonian formalism~\cite{Ryan}. The method is particularly useful since in some cases it makes it possible to find explicit solutions of Einstein's equations, and in general it allows a richer qualitative analysis of the dynamics of the models. Among the most studied models are the spatially homogeneous ones, in which the hypersurfaces $t=\text{const}$ are the orbits of a three-dimensional group of motions (isometries) acting simply transitively on them. There are nine types of such groups, and the corresponding models are known as Bianchi-type universes; one usually takes $N=1$ and $N^a=0$.

The general metric of these homogeneous models can be written as
\begin{equation}
\dd s^2=-\dd t^2+\qq_{ij}(t)\,\omega^i\omega^j\,,
\end{equation}
where the $\omega^i$ ($i=1,2,3$) are one-forms on $\Sig$ invariant under the group (for Bianchi type I they can be taken to be the $\dd x^i$). In abstract index notation, $g_{ab}=-\nabla_at\,\nabla_bt+\qq_{ij}(t)\,\omega^i{}_a\,\omega^j{}_b$.

The procedure described at the end of Sec.~\ref{sec:formal}, which consists in solving the constraints and choosing coordinate conditions in order to bring the action to canonical form (thus obtaining a true Hamiltonian), has been carried out in some of these cases: coordinates are defined as functions of $\qq_{ij}$ and $\pi^{ij}$, and the Hamiltonian is expressed in terms of them (see~\cite{Ryan}).

The three-geometries in (4.1) depend only on the time coordinate, so the problem has a finite number of degrees of freedom: the variables depend only on the time parameter, and not on all the coordinates as in the case of a general spacetime. In the language of superspace, the three-geometries of the homogeneous cosmological models form a finite-dimensional subset of it, which Misner christened \emph{minisuperspace}~\cite{Misner}.

In minisuperspace, where the configuration variables are at most the six components $\qq_{ij}(t)$, and where the supermetric has, as shown by DeWitt~\cite{DeWitt}, the signature $(-,+,+,+,+,+)$ (one ``timelike'' direction associated with the volume, and five ``spacelike'' ones associated with the shape of the three-geometry), one can ask whether there is a relation between Hamilton's equations, derived from the Hamiltonian, and the geodesic equation of the supermetric. From the constraint $\Hs=0$ defined in~\eqref{eq:H} we can define a Hamiltonian (whose value is zero) and write it as
\begin{equation}
H=\tfrac12\big(G^{AB}\pi_A\pi_B-\mathcal{U}\big)\,,
\end{equation}
where each pair of indices $(ij)$ has been replaced by a single index $A$: the configuration variables are $g^A\equiv \qq_{ij}$, the momenta are $\pi_A\equiv\pi^{ij}$, $G^{AB}$ stands for $G_{ijkl}$, and $\mathcal{U}(g)\equiv\sqrt{\qq}\;\sR$. Note that the form of the Hamiltonian (4.2) is similar to that of a free relativistic particle, $\tfrac12(g^{\mu\nu}p_\mu p_\nu+m^2)$, in a spacetime with metric $G_{AB}$ (which, as just noted, has Lorentzian signature). The only difference is that in (4.2) the role of the mass term is played by $-\mathcal{U}$, which may in general depend on the coordinates $g^A$.

Hamilton's equations for this Hamiltonian are
\begin{equation}
\begin{split}
\frac{\dd g^A}{\dd\lambda}&=\frac{\pd H}{\pd\pi_A}=G^{AB}\pi_B\,,\\
\frac{\dd\pi_A}{\dd\lambda}&=-\frac{\pd H}{\pd g^A}=-\tfrac12\,\pi_C\pi_D\,\pd_AG^{CD}+\tfrac12\,\pd_A\,\mathcal{U}\,,
\end{split}
\end{equation}
where $\pd_A\equiv\pd/\pd g^A$ and $\lambda$ is the parameter along the trajectory (with the normalization of (4.2), $\dd\lambda=2N\dd t$). From these equations we derive
\[
\frac{\dd^2g^A}{\dd\lambda^2}=\pd_CG^{AB}\,\frac{\dd g^C}{\dd\lambda}\,\pi_B+G^{AB}\Big(-\tfrac12\,\pi_D\pi_E\,\pd_BG^{DE}+\tfrac12\,\pd_B\,\mathcal{U}\Big)\,,
\]
which can be rewritten as
\[
\frac{\dd^2g^A}{\dd\lambda^2}+\tfrac12\,G^{AB}\big(\pd_CG_{BD}+\pd_DG_{BC}-\pd_BG_{CD}\big)\frac{\dd g^C}{\dd\lambda}\frac{\dd g^D}{\dd\lambda}=\tfrac12\,G^{AB}\,\pd_B\,\mathcal{U}\,.
\]
Therefore,
\begin{equation}
\frac{\dd^2g^A}{\dd\lambda^2}+\Gamma^A{}_{CD}\,\frac{\dd g^C}{\dd\lambda}\frac{\dd g^D}{\dd\lambda}=\tfrac12\,\pd^A\,\mathcal{U}\,,
\end{equation}
where $\Gamma^A{}_{CD}$ are the Christoffel symbols of $G_{AB}$ and $\pd^A\equiv G^{AB}\pd_B$; in addition, the constraint $H=0$ fixes the normalization of the tangent vector, $G_{AB}\,\frac{\dd g^A}{\dd\lambda}\frac{\dd g^B}{\dd\lambda}=\mathcal{U}$. That is, we obtain the geodesic equation of the metric $G_{AB}$ with an extra term. This equation makes it possible to relate, in a certain sense, Hamilton's equations to the geodesic equation, to which a potential has been added that is responsible for an extra force. If this term were somehow negligible, the cosmological model would behave essentially like a free particle in a curved spacetime.

\section{Developments since 1991}\label{sec:developments}

The applications described in Sec.~\ref{sec:applications} reflect the state of the field in 1991. In this section, which is new in the present edition, we sketch some of the main developments since then that rest directly on the 3+1 decomposition and on the Hamiltonian formulation described above. The selection is partial and the presentation brief; wherever possible we refer to reviews and textbooks rather than to the original literature.

\subsection{Numerical Relativity}\label{sec:NR}

That the Cauchy problem of general relativity is well posed was established long before the work described here: data $(\qq_{ab},K_{ab})$ satisfying the constraints have a development satisfying Einstein's equations~\cite{FouresBruhat}, and a unique maximal globally hyperbolic one~\cite{ChoquetBruhatGeroch}. The 3+1 formalism turns this result into a computational strategy: solve the constraints (3.81)--(3.82) on an initial hypersurface, choose the lapse and the shift, and integrate the evolution equations. In 1991 this programme was in its early stages~\cite{Matzner}; today the simulation of the coalescence of two black holes is routine. Three ingredients made this possible.

\paragraph{Initial data.} The constraints are four equations for the twelve components of $(\qq_{ab},K_{ab})$, so part of the data must be chosen freely. In the conformal method of Lichnerowicz and York~\cite{Lichnerowicz,York79} one writes $\qq_{ab}=\psi^4\tilde \qq_{ab}$ and decomposes $K_{ab}$ into its trace and its trace-free part. For maximal ($K=0$), conformally flat data with $K_{ab}=\psi^{-2}\tilde A_{ab}$, the momentum constraint becomes the linear equation $\pd_j\tilde A^{ij}=0$ for the trace-free tensor $\tilde A_{ij}$ (indices raised with $\delta^{ij}$), and the Hamiltonian constraint becomes the non-linear elliptic equation
\[
\delta^{ij}\pd_i\pd_j\psi+\tfrac18\,\psi^{-7}\tilde A_{ij}\tilde A^{ij}=0\,,
\]
which reduces to (3.100) for time-symmetric data. Bowen and York~\cite{BowenYork} gave explicit solutions $\tilde A_{ij}$ carrying arbitrary linear and angular momentum, and the puncture method of Brandt and Brügmann~\cite{BrandtBrugmann} provides a convenient way of solving the resulting equation for $\psi$ for several black holes.

\paragraph{Formulation of the evolution equations.} With prescribed lapse and shift, the system (3.70), (3.87) is only weakly hyperbolic~\cite{SarbachTiglio}. Its Cauchy problem is therefore not well posed in general, and numerical evolutions based on it tend to develop instabilities and to violate the constraints. The lesson, already visible in the comparison between the ADM and York systems in Sec.~\ref{sec:einstein-3+1}, is that systems which agree on the constraint surface can behave very differently off it. This led to several reformulations. In the BSSN formulation~\cite{ShibataNakamura,BaumgarteShapiro98} the conformal factor, the conformal metric of unit determinant, the trace $K$, the conformal trace-free part of $K_{ab}$ and the contracted conformal Christoffel symbols are evolved as independent variables, and multiples of the constraints are added to the evolution equations. The generalized harmonic formulation~\cite{Friedrich85,Pretorius05} writes Einstein's equations as wave equations for the metric components in a harmonic-type gauge with source functions. The Z4 family~\cite{BonaZ4,BernuzziHilditch,AlicCCZ4} introduces an additional vector field that propagates, and allows one to damp, constraint violations. With suitable gauge conditions all of these are strongly hyperbolic.

\paragraph{Gauge conditions.} The lapse and the shift are free (Sec.~\ref{sec:3+1}), and their choice is crucial in practice: the slicing must avoid the singularities inside black holes, and the coordinates must remain regular during the evolution. Maximal slicing, $K=0$, requires by the trace of (3.87) that the lapse satisfy the elliptic equation $\D_a\D^aN=N\,K_{ab}K^{ab}$ in vacuum, and it avoids singularities by making the lapse collapse in strong-field regions. In practice one uses hyperbolic slicings of the Bona--Massó family~\cite{BonaMasso}, in particular the ``1+log'' condition $(\pd_t-N^a\pd_a)N=2NK$,\footnote{With our sign convention for $K_{ab}$. In most of the numerical-relativity literature, which uses the opposite sign, this reads $(\pd_t-\beta^i\pd_i)\alpha=-2\alpha K$, with $\alpha$ the lapse and $\beta^i$ the shift.} combined with the ``Gamma-driver'' shift conditions~\cite{Alcubierre03}.

\medskip
In 2005 Pretorius~\cite{Pretorius05}, using the generalized harmonic formulation with black-hole excision, obtained the first evolution of a binary black hole through several orbits, the merger and the ringdown of the final black hole. A few months later, two groups~\cite{Campanelli06,Baker06} achieved the same with the BSSN formulation, 1+log slicing and Gamma-driver shift, letting the punctures move across the grid (the ``moving puncture'' method). Since then such simulations have become a standard tool, and waveform models calibrated against them were used in the analysis of the first direct detection of gravitational waves, GW150914, the signal of a binary black hole merger~\cite{GW150914}. Modern accounts of the subject can be found in~\cite{AlcubierreBook,BaumgarteShapiroBook,Gourgoulhon}.

\subsection{Canonical Quantum Gravity}

Of the two routes to quantization mentioned in Sec.~4.1, the second, Dirac's, has been followed most extensively. In the metric (Wheeler--DeWitt) form of this route~\cite{DeWitt}, beyond the problems of defining the operator products in $\hat{\Hs}$ at a point, one faces the \emph{problem of time}~\cite{Kuchar92,Isham93}. Since the Hamiltonian is a combination of constraints, the physical states do not evolve with respect to any external time, and time has to be recovered relationally, through the correlations between dynamical variables.

A new impulse came from Ashtekar's discovery of new canonical variables for general relativity~\cite{Ashtekar86}, in which the configuration variable is a connection. In the real formulation of Barbero~\cite{Barbero95}, the variables are the densitized triad $E^a{}_i$, with $E^a{}_iE^{bi}=\qq\,\qq^{ab}$, and the connection
\[
A_a{}^i=\Gamma_a{}^i+\gamma\,K_a{}^i\,,
\]
where $\Gamma_a{}^i$ is the spin connection of the triad, $K_a{}^i$ is the extrinsic curvature with one index converted into an internal index with the triad, and $\gamma$ is the Barbero--Immirzi parameter~\cite{Immirzi97}. The phase space is that of an $SU(2)$ gauge theory: besides the momentum and Hamiltonian constraints there is a Gauss constraint, $\D_aE^a{}_i=0$, generating internal rotations of the triad. Reducing by it gives back the ADM phase space of Sec.~\ref{sec:hamiltonian}.

Loop quantum gravity is the quantization of this formulation in which the basic variables are the holonomies of the connection along curves and the fluxes of the triad across surfaces~\cite{AshtekarLewandowski04,RovelliBook,ThiemannBook,AshtekarPullin}. Its kinematical Hilbert space is spanned by spin-network states, and it carries the unique representation of the holonomy--flux algebra that is invariant under spatial diffeomorphisms~\cite{LOST,Fleischhack}. Operators representing the area of surfaces and the volume of regions have discrete spectra~\cite{RovelliSmolin,AshtekarLewandowski97}; for instance, the area eigenvalues are $8\pi\gamma\,\ell_{\rm P}^2\sum_e\sqrt{j_e(j_e+1)}$, where $\ell_{\rm P}^2=G\hbar$ and the half-integers $j_e$ label the edges of the spin network crossing the surface. The momentum constraint is solved by averaging over diffeomorphisms. The Hamiltonian constraint has been defined as an operator~\cite{Thiemann98}, but the dynamics of the theory, and in particular the fate of the algebra (3.93) in the quantum theory, remain the subject of active research, also in the covariant ``spin foam'' formulation~\cite{EPRL}.

\subsection{Cosmology}

Two developments since 1991 build directly on the Hamiltonian treatment of homogeneous cosmologies described in Sec.~4.3.

The first concerns the approach to cosmological singularities. Misner's picture of the Mixmaster (Bianchi type IX) universe as a particle bouncing off potential walls in minisuperspace~\cite{Misner69}, and the oscillatory behaviour found by Belinskii, Khalatnikov and Lifshitz~\cite{BKL}, were extended by Damour, Henneaux and Nicolai~\cite{DHN} to generic, inhomogeneous solutions. Near a spacelike singularity the dynamics decouples point by point and becomes that of a ``cosmological billiard'': the particle moves freely in the space of logarithmic scale factors, with the Lorentzian metric inherited from the DeWitt supermetric, and is reflected by sharp walls. Equation~(4.4) is the smooth precursor of this picture. For vacuum gravity in four dimensions the billiard is chaotic, its walls being those of the Weyl chamber of a hyperbolic Kac--Moody algebra.

The second is loop quantum cosmology~\cite{Bojowald01,BojowaldLRR,AshtekarSingh}, the application of the methods of Sec.~5.2 to cosmological minisuperspaces. In the flat Friedmann model with a massless scalar field, which the Lorentzian signature of the minisuperspace metric allows one to use as an internal clock (Exercise~A.12), the quantum Hamiltonian constraint becomes a difference equation in the volume. States that are sharply peaked on a classical trajectory at late times follow it until the energy density reaches a critical value of the order of the Planck density, where the big bang is replaced by a quantum bounce~\cite{APS06a,APS06b}. Their dynamics is well approximated by the effective Friedmann equation $H^2=\frac{8\pi G}{3}\rho\,(1-\rho/\rho_c)$, with $\rho_c\approx0.41\,\rho_{\rm Pl}$ for the usual value of $\gamma$. The extent to which these results depend on the choices made in the quantization is discussed in~\cite{CorichiSingh}.

\subsection{Geometrodynamics Regained and the Covariant Phase Space}\label{sec:HKT}

The algebra (3.93) is not specific to general relativity. As shown by Teitelboim~\cite{Teitelboim73}, following ideas of Dirac~\cite{DiracLectures}, it is the algebra of deformations of a hypersurface embedded in spacetime: the structure functions $\qq^{ab}$ reflect the fact that two successive normal deformations, performed in opposite orders, differ by a tangential one determined by the metric of the hypersurface. Hojman, Kuchař and Teitelboim~\cite{HKT} reversed the argument, under the title ``geometrodynamics regained''. If the canonical variables are a three-metric and its momentum, the constraints are local and they represent the deformation algebra, then, under mild additional assumptions, the Hamiltonian is essentially fixed: it must be that of general relativity, with Newton's constant and the cosmological constant as the only free parameters. In this sense the dynamics of general relativity follows from the requirement that the evolution of the hypersurfaces be independent of the foliation used to describe it, which gives a precise meaning to the geometrodynamical picture of Sec.~4.2. That the algebra is not a Lie algebra means that it is not, by itself, the algebra of spacetime diffeomorphisms; Isham and Kuchař~\cite{IshamKuchar} showed how the latter can be represented canonically by extending the phase space with embedding variables, the coordinates of Kuchař's hyperspace.

A complementary development is the \emph{covariant phase space} formulation~\cite{CrnkovicWitten,LeeWald}. In it the phase space is the space of solutions of the field equations, with a symplectic structure constructed directly from the Lagrangian, without choosing a foliation; it is equivalent to the ADM phase space. For each vector field $\xi^a$ the corresponding Hamiltonian, when it exists, satisfies $\delta H_\xi=\oint(\delta Q_\xi-\iota_\xi\theta)$, where $Q_\xi$ is the Noether charge associated with $\xi^a$ and $\theta$ the symplectic potential~\cite{IyerWald}. For asymptotic time translations at spatial infinity this reproduces the ADM energy (3.99). Applied to stationary black holes it gives the first law of black-hole mechanics, with the entropy as a Noether charge, and at null infinity it yields the Bondi mass and its flux~\cite{WaldZoupas}.

\medskip
\noindent\emph{Further reading.} Appendix E of~\cite{Wald} and Chapter 4 of~\cite{Poisson} give concise treatments of the Lagrangian and Hamiltonian formulations; \cite{Gourgoulhon,AlcubierreBook,BaumgarteShapiroBook} cover the 3+1 formalism and numerical relativity; \cite{HenneauxTeitelboim} is a standard reference on constrained systems; and \cite{RovelliBook,ThiemannBook,KieferBook,AshtekarSingh} cover canonical quantum gravity and quantum cosmology.

\section*{Acknowledgements}

The authors thank Jemal Guven (ICN-UNAM) for valuable discussions and useful comments on the subject of this work, and Eduardo Nahmad-Achar for reviewing the manuscript. One of the authors thanks DGAPA-UNAM for the scholarship received during the development of this work.

\medskip
\noindent\emph{Note on the translation.} The English translation of \emph{Introducción al formalismo ADM} on which this revised edition is based was prepared by Ribhu Paul, who acknowledges the authorship of the original authors and preserved the content and the presentation of the original.

\appendix
\section{Exercises}\label{app:exercises}

These exercises, new in the present edition, accompany Secs.~\ref{sec:formal} and~\ref{sec:developments}. Unless otherwise stated we use the conventions of the text; when a mass appears we set $G=1$.

\paragraph{A.1 Hypersurfaces with a spacelike normal.} Repeat the derivation of Sec.~3.1 and of the Gauss formula (3.32) for a hypersurface whose unit normal satisfies $n^an_a=+1$, as for the spheres of Sec.~\ref{sec:3+1}. Show that the projector is $\qq^a{}_b=\delta^a{}_b-n^an_b$, that the normal component in (3.32) changes sign, and that the contracted Gauss equation becomes $q^{ac}q^{bd}R_{abcd}={}^{2}\!R-K^2+K_{ab}K^{ab}$ for a surface in a three-dimensional space. Verify it for a sphere of radius $r$ in $\mathbb{R}^3$, for which $K_{ab}=\qq_{ab}/r$ and ${}^{2}\!R=2/r^2$. (This is Gauss's \emph{theorema egregium}.)

\paragraph{A.2 Inverse metric and volume element.} Derive (3.41) and (3.42) from (3.40). Show that $g^{ab}=\qq^{ab}-n^an^b$ with $n^a=(t^a-N^a)/N$, and that $\det g_{\mu\nu}=-N^2\det \qq_{ij}$.

\paragraph{A.3 Hyperboloidal slices of Minkowski spacetime.} Inside the future light cone of the origin, Minkowski spacetime can be written as $\dd s^2=-\dd\tau^2+\tau^2\big(\dd\chi^2+\sinh^2\!\chi\,\dd\Omega^2\big)$ (the Milne form). For the slices $\tau=\text{const}$ show that $N=1$, $N^a=0$, $K_{ab}=\qq_{ab}/\tau$ and $\sR=-6/\tau^2$. Verify the Hamiltonian constraint (3.81) and the evolution equation (3.87) with $\fR_{ab}=0$. (Hint: the Ricci tensor of the unit hyperbolic space is $-2$ times its metric.)

\paragraph{A.4 Extrinsic curvature from the Lie derivative.} Show that $\qq_a{}^c\qq_b{}^d\,\Lie_{\vec N}\qq_{cd}=\D_aN_b+\D_bN_a$ for a spatial vector $N^a$, and that $\Lie_{Nn}\qq_{cd}=N\Lie_n\qq_{cd}$ when projected onto $\Sig$. Use these results to complete the derivation of (3.44).

\paragraph{A.5 The DeWitt supermetric.} Split a symmetric tensor as $h_{ab}=\tilde h_{ab}+\frac13h\,\qq_{ab}$, with $\qq^{ab}\tilde h_{ab}=0$. Show that $G^{abcd}h_{ab}h_{cd}=\sqrt{\qq}\,\big(\tilde h_{ab}\tilde h^{ab}-\frac23h^2\big)$, so that at each point the supermetric has signature $(-,+,+,+,+,+)$. Writing $\qq_{ab}=\qq^{1/3}r_{ab}$ with $\det r=1$ and $\tau=(32/3)^{1/2}\qq^{1/4}$, show that
\[
G^{abcd}\,\dd \qq_{ab}\,\dd \qq_{cd}=-\dd\tau^2+\tfrac{3}{32}\,\tau^2\,{\rm tr}\big(r^{-1}\dd r\,r^{-1}\dd r\big)\,,
\]
so that the five ``spacelike'' directions correspond to the unimodular metrics $r_{ab}$ and the ``timelike'' one to the volume.

\paragraph{A.6 Variation of the scalar curvature.} Show that $\delta\,\sR=-\sR^{ab}\delta \qq_{ab}+\D^a\D^b\delta \qq_{ab}-\D^c\D_c(\qq^{ab}\delta \qq_{ab})$. Use it to derive the terms of (3.85) that contain derivatives of the lapse, and the surface term in (3.97).

\paragraph{A.7 The bracket of two Hamiltonian constraints.} Using the result of A.6, show that in $\{H[N],H[M]\}$ only the terms with derivatives of $N$ and $M$ survive, and that they give $\int \dd^3x\,2\pi^{ab}\big(N\D_a\D_bM-M\D_a\D_bN\big)$. Integrate by parts to obtain the last line of (3.93). Verify the first two lines using (3.92).

\paragraph{A.8 Counting degrees of freedom in electromagnetism.} For the Maxwell field, the canonical variables are $A_a$ and the electric field $E^a$ on $\Sig$, and $A_0$ is a Lagrange multiplier enforcing the Gauss law $\D_aE^a=0$. Show that the Gauss law is first class and generates the gauge transformations $A_a\to A_a+\D_a\lambda$, and count the degrees of freedom as in (3.94). Compare with general relativity.

\paragraph{A.9 Schwarzschild and Brill--Lindquist data.} Verify that $\sR=-8\psi^{-5}\delta^{ij}\pd_i\pd_j\psi$ for $\qq_{ab}=\psi^4\delta_{ab}$. Show that $\psi=1+m/2r$ gives the $t=\text{const}$ slices of Schwarzschild in isotropic coordinates, and compute $E_{\rm ADM}$ from (3.99). Show that the surface $r=m/2$ is a minimal surface and that the map $r\to m^2/4r$ is an isometry of the slice that exchanges its two asymptotic regions. For Brill--Lindquist data, show that $E_{\rm ADM}=\sum_\alpha m_\alpha$.

\paragraph{A.10 Painlevé--Gullstrand slices.} The Schwarzschild metric can be written as $\dd s^2=-\dd t^2+\big(\dd r+\sqrt{2M/r}\,\dd t\big)^2+r^2\dd\Omega^2$. Show that the slices $t=\text{const}$ are flat, with $N=1$ and $N^r=\sqrt{2M/r}$. Show that $K_{rr}=\frac12\sqrt{2M}\,r^{-3/2}$ and $K_{\theta\theta}=K_{\phi\phi}/\sin^2\theta=-\sqrt{2Mr}$, and verify both constraints. Show that the ADM energy integral (3.99) vanishes for these slices, and explain why this does not contradict $M\neq0$. (Hint: compare the decay of $K_{ab}$ with (3.96).)

\paragraph{A.11 Maximal slicing.} Using (3.87) and (3.70), show that $(\pd_t-N^a\pd_a)K=\D_a\D^aN-N\,K_{ab}K^{ab}$ in vacuum on the constraint surface, and hence that $K=0$ is preserved if and only if $\D_a\D^aN=N\,K_{ab}K^{ab}$. Argue that a non-trivial lapse that tends to one at infinity must then be smaller than one in the interior, and relate this to the singularity-avoiding property of maximal slicing.

\paragraph{A.12 Flat Friedmann minisuperspace.} Take $\qq_{ij}=a(t)^2\delta_{ij}$ on a cell of coordinate volume $V_0$, with lapse $N(t)$ and vanishing shift, and add a homogeneous scalar field with Lagrangian density $-\frac12\sqrt{-g}\,g^{ab}\nabla_a\phi\nabla_b\phi$. Show that the total Lagrangian is $L=\frac{V_0}{2N}\big(-12\,a\dot a^2+a^3\dot\phi^2\big)$, that the minisuperspace metric is proportional to $-12\,a\,\dd a^2+a^3\dd\phi^2$, and that the Hamiltonian is $H=N\Big(-\frac{p_a^2}{24V_0a}+\frac{p_\phi^2}{2V_0a^3}\Big)$. Solve the constraint $H=0$ and show that, along classical solutions, $\phi$ is a monotonic function of $t$ and can be used as a clock (Sec.~5.3).


\end{document}